\documentclass[english,prd,superscriptaddress,nofootinbib,preprintnumbers,eqsecnum]{revtex4}
\usepackage{graphicx}
\usepackage{dcolumn}
\usepackage{bm}
\usepackage{latexsym}
\usepackage{amsfonts}
\usepackage{amssymb}
\usepackage{amsmath}
\usepackage{mathrsfs}
\usepackage{color}

\newcommand{\C}{{\cal C}}
\newcommand{\s}{{\cal S}}
\newcommand{\Z}{{\cal Z}}
\newcommand{\U}{{\cal U}}
\newcommand{\A}{{\cal A}}
\newcommand{\B}{{\cal B}}
\newcommand{\G}{{\cal G}}

\def\be{\begin{equation}}
\def\ee{\end{equation}}
\def\ba{\begin{eqnarray}}
\def\ea{\end{eqnarray}}

\begin{document}

\title{Effective field theory approach to modified gravity including \\ 
Horndeski theory and Ho\v{r}ava-Lifshitz gravity}

\author{Ryotaro Kase}
\affiliation{Department of Physics, Faculty of Science, Tokyo University of Science, 1-3,
Kagurazaka, Shinjuku, Tokyo 162-8601, Japan}

\author{Shinji Tsujikawa}
\affiliation{Department of Physics, Faculty of Science, Tokyo University of Science, 1-3,
Kagurazaka, Shinjuku, Tokyo 162-8601, Japan}

\date{\today}

\begin{abstract}

We review the effective field theory of modified gravity in which 
the Lagrangian involves three dimensional geometric quantities 
appearing in the 3+1 decomposition of space-time. 
On the flat isotropic cosmological background we expand a general 
action up to second order in the perturbations of geometric scalars, 
by taking into account spatial derivatives higher than two. 
Our analysis covers a wide range of gravitational theories-- including 
Horndeski theory/its recent generalizations and the 
projectable/non-projectable versions of Ho\v{r}ava-Lifshitz gravity. 
We derive the equations of motion for linear cosmological 
perturbations and apply them to the calculations of inflationary 
power spectra as well as the dark energy dynamics in 
Galileon theories. We also show that our general results conveniently 
recover stability conditions of Ho\v{r}ava-Lifshitz gravity already 
derived in the literature.

\end{abstract}

\maketitle




\section{Introduction}
\label{intro} 

The cosmic acceleration in the early Universe (dubbed inflation) 
was originally proposed to address flatness and horizon problems of 
the big-bang cosmology \cite{oldinf}.
Moreover, the inflationary paradigm has been strongly 
supported from the observations of Cosmic Microwave 
Background (CMB) temperature anisotropies \cite{WMAP9,Planck}. 
A key property of inflation is the existence of a scalar degree of freedom 
responsible for both the accelerated expansion at the background level 
and the generation of primordial density perturbations from 
quantum fluctuations \cite{oldper}.
Not only a canonical scalar field with a nearly flat potential \cite{newinf,chaotic} 
but also a gravitational scalar degree of freedom arising from
violation of the gauge symmetry of General Relativity 
(GR)--like Starobinsky inflation \cite{fRearly}-- can play such crucial roles.

In 1998, another late-time cosmic acceleration was discovered from the 
observations of type Ia supernovae (SN Ia) in high redshifts \cite{SNIa}.
The simplest origin for this present-day acceleration is the 
cosmological constant, but the vacuum 
energy appearing in particle physics is vastly larger than 
the observed energy scale of dark energy \cite{Weinbergco}. 
There is a possibility that the accelerated expansion 
of the Universe is driven by a scalar field \cite{quinpapers,kessence} 
or some modification of 
gravity \cite{DGP,fR1,fR2,scaten,Nicolis,Galileon,massive}. 
The modification of gravity is usually associated with 
the propagation of a scalar degree of freedom coupled 
to non-relativistic matter (see Refs.~\cite{moreview} for reviews).

The effective field theory (EFT) of cosmological perturbations 
is a powerful framework to deal with the low-energy degree of freedom 
of inflation and dark energy in a systematic and 
unified way \cite{Weinberg}-\cite{Gleyzes14}. 
This approach is based on the expansion of a general four-dimensional action 
about the flat Friedmann-Lema\^{i}tre-Robertson-Walker (FLRW) 
background in terms of the perturbations of three-dimensional geometric 
scalar quantities appearing in the 3+1 Arnowitt-Deser-Misner (ADM) \cite{ADM}
decomposition of space-time\footnote{On the spherically symmetric and static 
background there is another singled-out radial direction in addition to 
the temporal direction. Even in such cases
it is possible to construct the EFT of modified gravity \cite{KLT} by employing the 
2+1+1 decomposition of space-time developed in Refs.~\cite{radial}.}.
Such geometric scalars involve the traces and squares of the 
extrinsic curvature $K_{\mu \nu}$ and the three-dimensional 
intrinsic curvature ${\cal R}_{\mu \nu} \equiv {}^{(3)}R_{\mu \nu}$ 
as well as the lapse function $N$. 
The Lagrangian generally depends on a scalar field $\phi$ 
and its kinetic energy $X$, but such dependence can be absorbed 
into the lapse dependence by choosing the so-called unitary 
gauge in which the field perturbation $\delta \phi$ vanishes.

The EFT formalism can incorporate a wide variety of modified 
gravitational theories known in the literature.
For example, Horndeski \cite{Horndeski} derived a four-dimensional 
action of most general single-field scalar-tensor theories with 
second-order equations of motion in generic space-time 
(see also Refs.~\cite{Deffayet,KYY,Char}). 
In the EFT approach time derivatives are of second order 
by construction, but there exist spatial derivatives higher 
than second order in general \cite{Bloom,Bloom2}. 
In Ref.~\cite{Piazza}, the conditions for the absence of such 
higher-order spatial derivatives have been derived by expanding 
the action up to second order in the cosmological perturbations of geometric 
scalars. In fact the Horndeski theory satisfies such conditions, 
so the resulting second-order Lagrangian is simply expressed 
by the sum of time and spatial derivatives 
$\dot{\zeta}^2$ and $(\partial \zeta)^2$ 
of curvature perturbations $\zeta$ with time-dependent 
coefficients. Even on the spherically symmetric and static background 
it is possible to encompass the Horndeski theory by using  
geometric scalars appearing in the 2+1+1 ADM formalism \cite{KLT}.

While the Horndeski theory is Lorentz-invariant, the EFT approach 
can also cover Lorentz-violating theories such as 
Ho\v{r}ava-Lifshitz gravity \cite{Horava}. 
In Ho\v{r}ava-Lifshitz gravity, Lorentz invariance is explicitly broken to realize 
an anisotropic scaling in time $t$. 
This anisotropic scaling was introduced to suppress non-linear gravitational 
interactions in the ultraviolet (UV) regime. 
Ho\v{r}ava-Lifshitz gravity is based on the ADM formalism 
with a kinetic Lagrangian $L_K$ constructed from scalars associated with 
$K_{\mu \nu}$ and a ``potential'' term $L_{\cal V}$ constructed from scalars associated 
with ${\cal R}_{\mu \nu}$ and its spatial derivatives.
The presence of six spatial derivatives such as 
$\nabla_i{\cal R}_{jk} \nabla^i {\cal R}^{jk}$ allows the $z=3$ 
scaling characterized by the transformation $t \to c^3 t$ and $x^i \to cx^i$ 
($c$ is a constant and $x^i$ are spatial coordinates with $i=1,2,3$), 
in which case the theory is power-counting renormalizable. 

The original version of Ho\v{r}ava-Lifshitz gravity satisfies the so-called 
projectability condition under which the lapse $N$ is a function of time 
$t$ alone \cite{Horava}. 
This is consistent with the foliation-preserving diffeomorphism
characterized by $t \to t+f(t)$ and $x^i \to x^i+\xi^i (t,x^i)$. 
However, the theory is plagued by the problems of Laplacian instabilities 
and strong couplings even in the deep infrared (IR) 
regime \cite{strong,strong2}. 
These problems can be alleviated in the non-projectable version of 
Ho\v{r}ava-Lifshitz gravity where the lapse $N$ depends upon 
both $t$ and $x^i$ \cite{Sergey}. 
In this case the so-called acceleration vector 
$a_i=\nabla_i \ln N$, where $\nabla_i$ is the covariant derivative 
with respect to the three-dimensional spatial metric, does not vanish.
Scalar quantities constructed from $a_i$ allows a possibility of 
making the theory healthy \cite{Sergey,Blas}.

In this article we generalize the EFT approach of Ref.~\cite{Piazza} 
in such a way that the formalism can accommodate the cases 
in which spatial derivatives higher than second order appearing in 
the projectable/non-projectable versions of Ho\v{r}ava-Lifshitz 
gravity are present. We start with the general action involving the 
dependence of higher-order spatial derivatives such as 
$\nabla_i{\cal R}_{jk} \nabla^i {\cal R}^{jk}$ and $a_i a^i$ 
as well as of geometric scalar quantities constructed from $N$, 
$K_{\mu \nu}$, and ${\cal R}_{\mu \nu}$. 
Note that a similar approach was taken in Ref.~\cite{Xian}, 
but the contributions of higher-order spatial derivative terms 
to the second-order action of cosmological perturbations
were not explicitly computed  
for scalar perturbations. 

While our article is prepared for Special Issue in International 
Journal of Modern Physics D, it includes some new findings 
with an extension of the formalism already developed in the literature.
Moreover, the EFT approach explained in this article 
will be useful for researchers who are interested in the systematic 
study of a wide variety of modified gravitational theories including 
Horndeski theory and Ho\v{r}ava-Lifshitz gravity.
We shall derive the equations of motion for the background and 
linear perturbations convenient for discussing the dynamics of 
both inflation and dark energy.
These results will be useful for the construction of theoretical 
consistent models of cosmic accelerations 
and for the unified/systematic description
to test for such models with numerous observational data.

This paper is organized as follows.

In Sec.~\ref{theorysec} we introduce geometric 
scalar quantities appearing in the ADM formalism and 
provide a general action that depends on such scalars. 

In Sec.~\ref{concretesec} several modified gravitational 
theories accommodated in our EFT approach are reviewed.
Such theories include Horndeski theory and its recent 
generalization \cite{Gleyzes} as well as Ho\v{r}ava-Lifshitz gravity.

In Sec.~\ref{persec} we expand the action up to second order in 
cosmological perturbations and derive equations of 
motion for the background and the perturbations. 
We also derive conditions for the absence of spatial derivatives 
higher than second order and conditions for avoiding ghosts 
and Laplacian instabilities.

In Sec.~\ref{apsec1} we apply our general results to the cosmology 
of Horndeski theory and its recent generalization.
We derive the primordial power spectrum of curvature perturbations
generated during inflation driven by a single scalar degree of freedom.
We also study the dynamics of dark energy in the presence of 
an additional matter and clarify how Horndeski theory and its 
generalization can be distinguished from each other 
at the level of perturbations.

In Sec.~\ref{apsec2} we apply our formalism to both the projectable and 
non-projectable versions of Ho\v{r}ava-Lifshitz gravity and show how 
the perturbation equations of motion already derived both on the Minkowski 
and FLRW backgrounds can be reproduced 
in our general framework.
 
Sec.~\ref{consec} is devoted to conclusions.

Throughout the paper, Greek and Latin indices denote components in space-time 
and in a three-dimensional space-adapted basis, respectively.  
A dot represents a derivative with respect to the cosmic time $t$.
We use the metric signature $(-,+,+,+)$ and units 
$c=\hbar=k_B=1$.


\section{The general EFT action of modified gravity}
\label{theorysec} 

The EFT of cosmological perturbations is based upon 
the 3+1 decomposition of space-time described by 
the line element \cite{ADM}
\be
ds^{2}=g_{\mu \nu }dx^{\mu }dx^{\nu}
=-N^{2}dt^{2}+h_{ij}(dx^{i}+N^{i}dt)(dx^{j}+N^{j}dt)\,,  
\label{ADMmetric}
\ee
where $N$ is the lapse function, $N^i$ is the shift vector, and
$h_{ij}$ is the three-dimensional spatial metric. 
The components of the four-dimensional metric $g_{\mu \nu}$ 
can be expressed as $g_{00}=-N^2+N^iN_i$, $g_{0i}=g_{i0}=N_i$, 
and $g_{ij}=h_{ij}$. 

We define a unit vector orthogonal to the constant $t$ hypersurfaces 
$\Sigma_t$, as $n_{\mu}=-Nt_{;\mu}=(-N,0,0,0)$. 
The induced metric $h_{\mu \nu}$ on $\Sigma_t$ can be 
expressed in the form $h_{\mu \nu}=g_{\mu \nu}
+n_{\mu}n_{\nu}$. Since the unit vector $n^{\mu}=(1/N,-N^{i}/N)$ 
obeys the relation $n_{\mu}n^{\mu}=-1$, it follows that 
$n^{\mu} h_{\mu \nu}=0$.

The covariant derivative of the vector $n_{\nu}$ with respect 
to $g_{\mu \nu}$ (denoted as $n_{\nu;\lambda}$) 
corresponds to the directional derivative with 
respect to a parallel transport of $n_{\nu}$ in four-dimensional 
space-time. We define the extrinsic curvature as a projection 
of the tensor $n_{\nu;\lambda}$ on $\Sigma_t$, as
\be
K_{\mu \nu}=h^{\lambda}_{\mu} n_{\nu;\lambda}
=n_{\nu;\mu}+n_{\mu} a_{\nu}\,,
\label{Kmunu}
\ee
where $a_{\nu} \equiv n^{\lambda}n_{\nu;\lambda}$ is dubbed 
the acceleration. The extrinsic curvature is a three-dimensional 
quantity (satisfying the relation $n^{\mu}K_{\mu \nu}=0$) that 
depends on the particular embedding of $\Sigma_t$.

The acceleration $a_{\nu}$ characterizes the difference between $K_{\mu \nu}$ 
(the change induced by the parallel transport of $n_{\nu}$ on $\Sigma_t$) and 
$n_{\nu;\mu}$ (the change induced by the parallel transport of 
$n_{\nu}$ in four dimensions).
Substituting the relation $n_{\nu}=-Nt_{;\nu}$ into 
$a_{\nu}=n^{\lambda}n_{\nu;\lambda}$ and using the 
properties $t_{;\nu \lambda}=t_{;\lambda \nu}$,
$n^{\lambda}n_{\lambda;\nu}=0$, 
$h^{\lambda}_{\nu}=\delta^{\lambda}_{\nu}+n_{\nu}n^{\lambda}$, 
the acceleration can be written in the form
\be
a_{\nu}=\frac{h^{\lambda}_{\nu} N_{;\lambda}}{N} 
=\nabla_{\nu} \ln N\,,
\label{acc}
\ee
where $\nabla_{\nu}$ represents the covariant derivative 
with respect to the three-dimensional metric $h_{ij}$. 
If $N$ is a function of time $t$ alone, then 
the acceleration vanishes.
In general, however, $N$ depends on the spatial coordinate 
and hence $a_{\nu} \neq 0$.

{}From Eq.~(\ref{Kmunu}) the extrinsic curvature can be expressed
as $K_{ij}=[n_{i;j}+n_{j;i}+n^{\mu}(n_{i}n_{j})_{;\mu}]/2$, 
where we used the symmetric property $K_{ij}=K_{ji}$. 
Taking note of the relations $n_{i;j}=h_{i\alpha}{n^{\alpha}}_{;j}$, 
${n^{\alpha}}_{;j}=\partial_{j}n^{\alpha}+\Gamma^{\alpha}_{\mu j}n^{\mu}$ 
and $h_{ij;\mu}=\partial_{\mu}h_{ij}-\Gamma^{\nu}_{i \mu} h_{\nu j}
-\Gamma^{\nu}_{j \mu}h_{i\nu}$, 
where $\Gamma^{\alpha}_{\mu j}$ is the Christoffel symbol 
with respect to $g_{\mu \nu}$, it follows that
\be
K_{ij}=\frac12 \left( \frac{1}{N} \partial_t h_{ij}
+n^{k} \partial_{k} h_{ij}+h_{j \alpha} \partial_{i} n^{\alpha}
+h_{i \alpha} \partial_{j} n^{\alpha} \right)\,.
\label{Kijm}
\ee
Using the three-dimensional covariant derivative 
$\nabla_iN_{j}=\partial_{i}N_{j}-{}^{(3)}\Gamma^{k}_{ij}N_{k}$, 
where ${}^{(3)}\Gamma^{k}_{ij}$ is the Christoffel symbol 
associated with $h_{ij}$, the extrinsic curvature (\ref{Kijm})
is simply expressed as 
\be
K_{ij}=\frac{1}{2N} \left( \partial_t h_{ij}
-\nabla_i N_{j}-\nabla_j N_{i} \right)\,.
\label{Kijm2}
\ee
Hence $K_{ij}$ is a covariant quantity that involves 
a time derivative of the three-dimensional metric $h_{ij}$.

The kinetic scalar quantities that can appear in the action of 
general modified gravitational theories are derived by 
taking the trace of $K_{\mu \nu}$ and by squaring 
$K_{\mu \nu}$, as 
\be
K \equiv {K^{\mu}}_{\mu}\,,\qquad
\s \equiv K_{\mu \nu} K^{\mu \nu}\,.
\ee

The internal geometry of $\Sigma_t$ is quantified
by the three-dimensional Ricci tensor 
${\cal R}_{\mu \nu}={}^{(3)}R_{\mu \nu}$ 
(dubbed the intrinsic curvature). 
The scalar quantities constructed from 
${\cal R}_{\mu \nu}$ are given by 
\be
\mathcal{R} \equiv
{\mathcal{R}^{\mu}}_{\mu}\,,\qquad
\Z \equiv \mathcal{R}_{\mu \nu}
\mathcal{R}^{\mu \nu}\,.
\ee
From ${\cal R}_{\mu \nu}$ and $K^{\mu \nu}$ 
we can construct another scalar quantity: 
\be
\U \equiv \mathcal{R}_{\mu \nu}K^{\mu \nu}\,. 
\ee
Note that the three-dimensional Riemann tensor 
${}^{(3)}R_{\mu \nu \rho \sigma}$ gives rise to 
the quadratic combination 
${}^{(3)}R_{\mu \nu \rho \sigma}{}^{(3)}R^{\mu \nu \rho \sigma}$.
In three dimensions, however, the Riemann tensor 
can be expressed in terms of the Ricci tensor and scalar, 
so we do not need to consider such a combination.

We also allow for the existence of scalar quantities that give rise to 
spatial derivatives higher than second order in the equations 
of motion:
\be
{\cal Z}_1 \equiv \nabla_i {\cal R} \nabla^i {\cal R}\,,\qquad
{\cal Z}_2 \equiv \nabla_i {\cal R}_{jk} \nabla^i {\cal R}^{jk}\,. 
\label{Zdef}
\ee
Other terms such as 
${\cal R}^{j}_{i} {\cal R}^{k}_{j}  {\cal R}^{i}_{k}$ and 
${\cal R} {\cal R}^{j}_{i} {\cal R}^{i}_{j}$ can be taken into account, 
but they are irrelevant to scalar linear perturbations on the flat 
FLRW background studied in Sec.~\ref{persec}. 
We do not include the terms with more than six spatial 
derivatives, as we are interested in the application to 
Ho\v{r}ava-Lifshitz gravity.
Provided that the theory is power-counting renormalizable by an anisotropic 
scaling with $z=3$ (discussed in Sec.~\ref{Horavasec}), 
such higher-order terms are not generated by quantum corrections.

In the original version of Ho\v{r}ava-Lifshitz gravity \cite{Horava}, 
the lapse $N$ is assumed to be a function of time $t$ alone (which is called 
the projectability condition). This reflects the fact that the space-time 
foliation is preserved by the space-independent reparametrization 
$t \to t'(t)$. In this case, the acceleration $a_{\nu}$ of Eq.~(\ref{acc}) 
vanishes and hence $K_{\mu \nu}=n_{\nu;\mu}$. 
One can extend the original Ho\v{r}ava-Lifshitz theory such that the lapse depends 
on the spatial coordinate $x^i$ ($i=1,2,3$) as well as time $t$ and 
that the acceleration $a_{\nu}$ is included in the action \cite{Sergey}.
In this non-projectable version 
the following scalar combinations can be taken into account:
\be
\alpha_1 \equiv a_i a^i\,,\qquad
\alpha_2 \equiv a_i \Delta a^i\,,\qquad
\alpha_3 \equiv {\cal R} \nabla_i a^i\,,\qquad
\alpha_4 \equiv a_i \Delta^2 a^i\,,\qquad
\alpha_5 \equiv \Delta {\cal R} \nabla_i a^i\,,
\label{aldef}
\ee
where $\Delta \equiv \nabla_i \nabla^i$. 
Again we do not include the terms irrelevant to the dynamics 
of linear scalar perturbations on the flat FLRW background
(such as $(a_i a^i)^2$ and $a_i a_j {\cal R}^{ij}$).

The action of general modified gravitational theories that depends on 
the above mentioned scalar quantities is given by 
\be
S=\int d^4 x \sqrt{-g}\,L \left( N, K, \s, {\cal R}, {\cal Z}, {\cal U}, 
{\cal Z}_1, {\cal Z}_2, \alpha_1, \cdots, \alpha_5; t \right)\,,
\label{action}
\ee
where $g$ is a determinant of the metric $g_{\mu \nu}$ and 
$L$ is a Lagrangian. 
The dependence of the lapse $N$ and the time 
$t$ is included for the reason explained in Sec.~\ref{Hornsec}.
Expanding the action (\ref{action}) up to second order in cosmological 
perturbations about the flat FLRW background, we obtain the equations 
of motion for the background and linear perturbations.
Before doing so, we shall review the theories that belong to the action 
(\ref{action}).


\section{Concrete theories accommodated in the EFT framework}
\label{concretesec} 

The EFT formalism can deal with a wide variety of gravitational 
theories-- including (i) Horndeski theory \cite{Horndeski} 
and its generalization \cite{Gleyzes}, 
and (ii) Ho\v{r}ava-Lifshitz gravity \cite{Horava}.
In this section we discuss explicit dependence of the Lagrangian $L$
on the geometric scalar quantities introduced in Sec.~\ref{theorysec}.

\subsection{Horndeski theory and its generalization}
\label{Hornsec}

The Lagrangian of most general scalar-tensor theories with second-order 
equations of motion was first derived by Horndeski in 1973 \cite{Horndeski}. 
In four dimensions, Horndeski theory is characterized by the
Lagrangian \cite{Horndeski,Deffayet,Char,KYY}
\begin{equation}
L=\sum_{i=2}^{5} L_{i}\,,
\label{Lagsum}
\end{equation}
with
\begin{eqnarray}
L_{2} & = & G_2(\phi,X),\label{eachlag2}\\
L_{3} & = &  G_{3}(\phi,X)\square\phi,\label{eachlag3}\\
L_{4} & = & G_{4}(\phi,X)\, R-2G_{4,X}(\phi,X)\left[ (\square \phi)^{2}
-\phi^{;\mu \nu }\phi _{;\mu \nu} \right] \,, \label{L4lag}\\
L_{5} &=&G_{5}(\phi,X)G_{\mu \nu }\phi ^{;\mu \nu}
+\frac{1}{3}G_{5,X}(\phi,X)
\left[ (\square \phi )^{3}-3(\square \phi )\,\phi _{;\mu \nu }\phi ^{;\mu
\nu }+2\phi _{;\mu \nu }\phi ^{;\mu \sigma }{\phi ^{;\nu}}_{;\sigma }
\right]\,, 
\label{L5lag}
\end{eqnarray}
where $\square \phi \equiv (g^{\mu \nu} \phi_{;\nu})_{;\mu}$, and
$G_{i}$ ($i=2,3,4,5$) are functions in terms of a scalar
field $\phi$ and its kinetic energy 
$X=g^{\mu \nu}\partial_{\mu} \phi \partial_{\nu} \phi$
with the partial derivatives $G_{i,X} \equiv\partial G_{i}/\partial X$ 
and $G_{i,\phi} \equiv \partial G_{i}/\partial \phi$. 
$R$ and $G_{\mu\nu}$ are the Ricci scalar and 
the Einstein tensor in four dimensions, respectively.

Horndeski theory covers a wide variety of gravitational theories 
with a single scalar degree of freedom.
First of all, the k-essence scalar field \cite{kinf,kessence} in 
the framework of GR is described by the functions 
$G_2=P(\phi,X), G_3=0, G_4=M_{\rm pl}^2/2, 
G_5=0$, where $M_{\rm pl}$ is the reduced Planck mass.
The canonical scalar field with a potential $V(\phi)$ corresponds to 
a particular function $G_2=-X/2-V(\phi)$. 

Brans-Dicke (BD) theory \cite{Brans} with a potential $V(\phi)$ is 
characterized by the functions 
$G_2=-M_{\rm pl}\omega_{\rm BD} X/(2\phi)-V(\phi)$, 
$G_3=0$, $G_4=M_{\rm pl}\phi/2$, $G_5=0$, where 
$\omega_{\rm BD}$ is the BD parameter. 
The metric $f(R)$ gravity \cite{fR1,fR2} and dilaton gravity \cite{dilaton} 
correspond to the particular cases of BD theory with 
$\omega_{\rm BD}=0$ and $\omega_{\rm BD}=-1$, respectively.

The covariant Galileon \cite{Galileon} corresponds to the functions 
$G_2=\beta_2 X$, $G_3=\beta_3 X$, 
$G_4=M_{\rm pl}^2/2+\beta_4 X^2$, 
$G_5=\beta_5 X^2$, where $\beta_i$ ($i=2,3,4,5$) 
are constants.
A scalar field whose derivatives couple to 
the Einstein tensor in the form 
$G_{\mu \nu} \partial^{\mu} \phi \partial^{\nu} \phi$ \cite{Amendola93,Germani}
can be accommodated by the functions
$G_2=-X/2-V(\phi)$, $G_3=0$, $G_4=0$, $G_5=c \phi$, 
where $c$ is a constant and $V(\phi)$ is a field potential.
A proxy theory to massive gravity proposed in Ref.~\cite{Heisen}
corresponds to $G_2=0$, $G_3=0$, 
$G_4=M_{\rm pl}^2/2-M_{\rm pl}\phi-c_4X$, $G_5=c_5 X$, 
where $c_4$ and $c_5$ are constants.

The Lagrangian (\ref{Lagsum}) with (\ref{eachlag2})-(\ref{L5lag}) 
involves the dependence of $\phi$ and $X$, 
whereas the Lagrangian $L$ in the action (\ref{action})
does not explicitly depend on such a scalar field and its kinetic term.
However, if we choose the unitary gauge
\be
\phi=\phi(t)\,,
\ee
in which the perturbation $\delta \phi (t,x^i)$ on the flat FLRW 
background vanishes, the field kinetic term is expressed as
$X=-\dot{\phi}^2(t)/N^2$. 
Then, the $\phi$ and $X$ dependence can be interpreted as 
the $N$ and $t$ dependence appearing in the action (\ref{action}). 
In this way, it is possible to incorporate
Horndeski theory in the EFT formalism.

More explicitly, the Horndeski Lagrangian (\ref{Lagsum}) 
with (\ref{eachlag2})-(\ref{L5lag}) 
can be expressed in terms of the three-dimensional scalar quantities 
introduced in Sec.~\ref{theorysec} \cite{Piazza} 
(see also Refs.~\cite{Bloom2,Laszlo,Tsuji14}).  
In unitary gauge the constant-$\phi$ hypersurfaces 
coincide with $\Sigma_t$, so the unit vector orthogonal to 
those hypersurfaces is given by \cite{Piazza}
\be
n_{\mu}=-\gamma \phi_{;\mu}\,,\qquad
\gamma=\frac{1}{\sqrt{-X}}\,.
\label{nmu}
\ee

Employing the relation (\ref{Kmunu}) and taking the 
covariant derivative of Eq.~(\ref{nmu}), we obtain 
\be
\phi_{;\mu \nu} =-\frac{1}{\gamma}\left( K_{\mu \nu}
-n_{\mu }a_{\nu}-n_{\nu }a_{\mu}\right) 
+\frac{\gamma ^{2}}{2} 
\phi^{;\lambda}X_{;\lambda}n_{\mu }n_{\nu}\,,
\label{phimunu}
\ee
and hence 
\be
\square \phi=-\frac{1}{\gamma}K+
\frac{\phi^{;\lambda}X_{;\lambda}}{2X}\,.
\label{sqphi}
\ee
Then, the Lagrangian (\ref{eachlag3}) is expressed as 
$L_3=G_3[-K/\gamma+\phi^{;\lambda}X_{;\lambda}/(2X)]$, 
whose second term can be eliminated by introducing an auxiliary function 
$F_3(\phi,X)$ satisfying
\be
G_3=F_3+2X F_{3,X}\,.
\label{F3}
\ee
The contribution $F_3 \square \phi$ reduces to 
$-(F_{3,\phi} \phi_{;\lambda}+F_{3,X}X_{;\lambda})\phi^{;\lambda}$ 
up to a boundary term. The second term is cancelled by 
one of the terms in $2X F_{3,X} \square \phi$.
Then, the Lagrangian (\ref{eachlag3}) can be expressed as 
\be
L_3=2(-X)^{3/2}F_{3,X}K-XF_{3,\phi}\,.
\label{L3three}
\ee
Since $F_3$ depends on $\phi(t)$ and $X(t,N)=-\dot{\phi}^2(t)/N^2$, 
$L_3$ is a function of $N$, $K$, and $t$. 
The equations of motion following from the Lagrangian $L_3$ 
can be written in terms of $G_3$ and its derivatives with 
respect to $\phi$ and $X$, without containing 
the auxiliary function $F_3$ \cite{Piazza}.
 
Substituting Eqs.~(\ref{phimunu}) and (\ref{sqphi}) 
into Eq.~(\ref{L4lag}) and using the fact that 
$a_{\mu}=-h_{\mu}^{\nu}X_{;\nu}/(2X)$, 
the Lagrangian $L_4$ reads
\be
L_4=G_4R+2XG_{4,X} (K^2-{\cal S})
+2G_{4,X}X_{;\mu} (Kn^{\mu}-a^{\mu})\,.
\label{L4ex}
\ee
{}From the Gauss-Codazzi equations, the four-dimensional 
Ricci scalar $R$ is related to the three-dimensional 
Ricci scalar ${\cal R}$ according to 
\be
R={\cal R}-K^2+{\cal S}
+2 ( Kn^{\mu}-a^{\mu} )_{;\mu}\,.
\label{Rre}
\ee
On using Eq.~(\ref{Rre}) together with the relations 
$G_{4,X}X_{;\mu}=G_{4;\mu}+\gamma^{-1}G_{4,\phi}n_{\mu}$ 
and $n_{\mu}a^{\mu}=0$, Eq.~(\ref{L4ex}) reduces to  
\be
L_4=G_4{\cal R}+(2XG_{4,X}-G_4)(K^2-{\cal S})
-2\sqrt{-X}G_{4,\phi}K\,.
\label{L4three}
\ee

Similarly, the Lagrangian (\ref{L5lag}) can be expressed in terms 
of the three-dimensional quantities, as \cite{Piazza}
\be
L_{5} =
\frac{1}{2}XG_{5,\phi} (K^{2}-{\cal S})
-\frac13 (-X)^{3/2}G_{5,X}K_3
+\frac{1}{2}X(G_{5,\phi}-F_{5,\phi})\mathcal{R}
-\sqrt{-X}F_{5}\left( {\cal U}-\frac{1}{2}K\mathcal{R} 
\right) \,,
\label{L5three}
\ee
where $F_5(\phi,X)$ is an auxiliary function satisfying 
\be
G_{5,X}=\frac{F_5}{2X}+F_{5,X}\,,
\label{F5}
\ee
and 
\be
K_3 \equiv K^3-3KK_{ij}K^{ij}
+2K_{ij}K^{il}{K^j}_l\,.
\ee
Up to quadratic order in perturbations, the term $K_3$ 
is given by 
\be
K_3=3H \left( 2H^2-2KH+K^2-{\cal S} \right)+O(3)\,.
\ee

{}From Eqs.~(\ref{eachlag2}), (\ref{L3three}), (\ref{L4three}), and 
(\ref{L5three}) the total Lagrangian (\ref{Lagsum}) involves
the functions $\phi$, $X$, $K$, $K^2-{\cal S}$, ${\cal R}$, 
$K_3$, and ${\cal U}-K{\cal R}/2$. 
The dependence on $\phi$ and $X$ can be interpreted as 
that on $N$ and $t$. Then, the Horndeski Lagrangian 
is equivalent to 
\be
L=A_2(N,t)+A_3(N,t)K+A_4(N,t) (K^2-{\cal S})+B_4(N,t){\cal R}
+A_5(N,t) K_3+B_5(N,t) \left( {\cal U}-K {\cal R}/2 \right)\,,
\label{LH2}
\ee
where 
\ba
& & A_2=G_2-XF_{3,\phi}\,,\qquad 
A_3=2(-X)^{3/2}F_{3,X}-2\sqrt{-X}G_{4,\phi}\,,\qquad
A_4=2XG_{4,X}-G_4+XG_{5,\phi}/2\,,\nonumber \\
& & B_4=G_4+X(G_{5,\phi}-F_{5,\phi})/2\,,\qquad
A_5=-(-X)^{3/2}G_{5,X}/3\,,\qquad
B_5=-\sqrt{-X}F_{5}\,.
\label{AB}
\ea
The coefficients $A_4$ and $A_5$ 
are related to $B_4$ and $B_5$, as
\be
A_4=2XB_{4,X}-B_4\,,\qquad
A_5=-XB_{5,X}/3\,,
\label{ABcon}
\ee
under which the number of 6 independent functions 
reduces to 4.

Gleyzes, Langlois, Piazza, and Vernizzi (GLPV) \cite{Gleyzes} 
generalized Horndeski theory in such a way that the coefficients 
$A_4$, $A_5$, $B_4$, and $B_5$ 
are not necessarily related to each other.
Even in this case, the background and linear perturbation equations 
of motion about the flat FLRW background remain of second order 
with no additional scalar propagating degrees of freedom.
Taking the inverse procedure to that presented above, 
the GLPV Lagrangian (\ref{LH2}) can be expressed in terms 
of the scalar field $\phi$ and its covariant derivatives \cite{Gleyzes}.

The Lagrangians of both Horndeski theory and GLPV theory involve 
the dependence of $N, {\cal S}, K, {\cal R}, {\cal U}$, and $t$.

\subsection{Ho\v{r}ava-Lifshitz gravity}
\label{Horavasec}

The renormalization of GR is a difficult task because of the presence 
of non-linear graviton interactions.  
Ho\v{r}ava-Lifshitz gravity \cite{Horava} is an attempt to suppress 
such non-linear interactions in the UV regime 
by violating Lorentz symmetry of GR. 

In order to understand the basic idea of Ho\v{r}ava, 
we begin with standard field theory in Minkowski space-time (i.e., 
without including gravity) \cite{Mukoreview}. 
We consider the following 
anisotropic scaling 
\be
t \to c^zt\,,\qquad x \to cx\,,
\label{scaling}
\ee
where $c$ is an arbitrary number, and $z$ is a number 
dubbed dynamical critical exponent. 
Then, the action of a kinetic term of a scalar field 
$\varphi$ transforms as 
\be
\int dt d^3 x\, \frac12 \dot{\varphi}^2 \to 
c^{3-z+2s} \int dt d^3 x\, \frac12 \dot{\varphi}^2\,,
\ee
where we assumed the scaling $\varphi \to c^s\varphi$ for 
the field. The kinetic term is invariant under the condition 
\be
s=\frac{z-3}{2}\,.
\ee
When $z=3$, the scalar field is unchanged ($s=0$) under the 
anisotropic scaling (\ref{scaling}).
If we consider the $n$-th order interaction term $\varphi^n$, 
the corresponding action transforms as
\be
\int dt d^3 x\, \varphi^n \to 
c^{z+3+ns} \int dt d^3 x\, \varphi^n
\propto E^{-(z+3+ns)/z}  \int dt d^3 x\, \varphi^n\,,
\ee
where in the last proportionality we used the fact that the energy 
$E$ scales as $E \to c^{-z}E$. 
When $z=3$, the exponent $-(z+3+ns)/z$ is $-2$ for any $n$, so 
the non-linear interactions are power-counting renormalizable.
This power-counting renormalizability also holds for the anisotropic 
scaling with $z>3$.

The example of a scalar-field action realizing the invariance 
under the $z=3$ scaling is given by 
\be
S_{\rm UV}=\int dt d^3 x \left( \frac12 \dot{\varphi}^2
+\frac{\varphi \Delta^3 \varphi}{M^4} \right)\,, 
\label{SUV}
\ee
where $M$ is a constant having a dimension of mass. 
If we also take into account the Lagrangians 
$\varphi \Delta^2 \varphi$ and $\varphi \Delta \varphi$ obeying 
the $z=2$ and $z=1$ scalings, respectively, 
the resulting action is
\be
S=\int dt d^3 x \left( \frac12 \dot{\varphi}^2
+\frac{\varphi \Delta^3 \varphi}{M^4} 
+\frac{c_1 \varphi \Delta^2 \varphi}{M^2}+c_2^2 
\varphi \Delta \varphi
\right)\,, 
\label{Sfield}
\ee
where $c_1$ and $c_2$ are dimensionless constants.
In the UV region the third and fourth terms 
on the r.h.s. of Eq.~(\ref{Sfield}) are suppressed relative to 
the second one, so the action (\ref{Sfield}) reduces 
to (\ref{SUV}). This is the regime in which non-linear field 
interactions are suppressed due to the $z=3$ scaling. 
In the IR regime the fourth term on the r.h.s. 
of Eq.~(\ref{Sfield}) dominates over the second and third terms, 
so the resulting action $S_{\rm IR}=\int dt d^3 x ( \dot{\varphi}^2/2+c_2^2 
\varphi \Delta \varphi )$ is invariant under the $z=1$ scaling.

Ho\v{r}ava \cite{Horava} applied the above idea of anisotropic scaling to 
the construction of a power-counting renormalizable gravitational theory. 
Due to the privileged role of time, the theory should respect the symmetry 
under time reparametrization and time-dependent spatial diffeomorphism:
\be
t \to \tilde{t}(t)\,,\qquad 
x^i \to \tilde{x}^i (t, x^i)\,,\qquad (i=1,2,3),
\label{defeo}
\ee
under which Lorentz symmetry is explicitly broken. 
Since the time transformation is not spatially dependent, 
the foliation of space-time in terms of the hypersurfaces 
$\Sigma_t$ is always preserved.

Under the infinitesimal change, $t \to t+f(t)$ and 
$x^i \to x^i+\xi^i (t,x^i)$, the quantities $N$, $N_i$, and 
$h_{ij}$ appearing in the ADM metric (\ref{ADMmetric}) transform as
\ba
N &\to & N-\dot{f}N-f \dot{N}-\xi^i \partial_i N\,,\label{Ntrans}\\
N_i &\to& N_i-\dot{f}N_i-f\dot{N}_i-\dot{\xi}^j h_{ij}
-\nabla_i \xi^j N_j-\xi^j \nabla_j N_i\,,\\
h_{ij} &\to& h_{ij}-f\dot{h}_{ij}-h_{ik} \nabla_j \xi^k
-h_{jk} \nabla_i \xi^k\,.
\ea
If the lapse $N$ is a function of $t$ alone, the transformation 
(\ref{Ntrans}) induces only the time-dependent term $-\partial_t(fN)$.
Hence the condition $N=N(t)$ (dubbed projectability condition) is consistent 
with the foliation-preserving diffeomorphism.
In this case, the acceleration $a_i=\nabla_i \ln N$ vanishes.
We note, however, that the projectability condition is not mandatory 
and that we can consider a non-projectable version of the theory 
characterized by $N=N(t,x^i)$.

In GR, the Lagrangian without matter is simply given by 
$L_{\rm GR}=M_{\rm pl}^2 R/2$, where $M_{\rm pl}$ is 
related to the gravitational constant $G$ as 
$M_{\rm pl}^2=(8\pi G)^{-1}$. 
Using the relation (\ref{Rre}) and dropping a boundary term, 
the four-dimensional action of GR reads
\be
S_{\rm GR}= \int N dt \sqrt{h}\, d^3x\, L_{\rm GR}\,,\qquad
L_{\rm GR}=\frac{M_{\rm pl}^2}{2} \left( {\cal S}-K^2
+{\cal R} \right)\,,
\ee
where $h$ is a determinant of the metric $h_{ij}$.
Since the extrinsic curvature (\ref{Kijm2}) involves a time derivative of 
the metric $h_{ij}$, the two scalar quantities ${\cal S}=K_{\mu \nu}K^{\mu \nu}$ and 
$K^2=({K^{\mu}}_\mu)^2$ play the role of kinetic energies associated with 
the ``velocity'' $\partial_t h_{ij}$.  
In GR, only the combination ${\cal S}-K^2$ is allowed due to 
a gauge symmetry of the theory.

In Ho\v{r}ava-Lifshitz gravity, both ${\cal S}$ and $K^2$ are invariant 
under the foliation-preserving diffeomorphism (\ref{defeo}).
Hence the kinetic Lagrangian of this theory is given by  
\be
L_K=\frac{M_{\rm pl}^2}{2} \left( {\cal S}-\lambda K^2 \right)\,,
\label{LK}
\ee
where $\lambda$ is an arbitrary constant. 
GR corresponds to the case $\lambda=1$.

The three-dimensional Ricci scalar ${\cal R}$ involves 
the second derivatives of $h_{ij}$ with respect to the spatial 
coordinate $x^i$, whereas ${\cal S}$ and $K^2$ possess
the second time derivatives of $h_{ij}$.
Under the $z=1$ scaling, i.e., $t \to ct$ and $x^i \to c x^i$, 
the term ${\cal R}$ scales in the same way as 
${\cal S}$ and $K^2$. 
In order to realize the $z=3$ scaling we need to take into 
account the terms involving six spatial derivatives 
such as ${\cal Z}_1$ and ${\cal Z}_2$ in Eq.~(\ref{Zdef}).
In the non-projectable version of Ho\v{r}ava-Lifshitz gravity 
the acceleration $a_i=\nabla_i \ln N$ 
does not vanish, in which case the terms like 
$\alpha_4$ and $\alpha_5$ in Eq.~(\ref{aldef}) 
also exhibit the $z=3$ scaling. 
Then, the action invariant under the $z=3$ scaling is 
given by $S_{{\cal V}_3}=\int N dt \sqrt{h}\,d^3x\,L_{{\cal V}_3}$, 
with the Lagrangian
\be
L_{{\cal V}_3}=-\frac{1}{2M_{\rm pl}^2}
\left( g_4 {\cal Z}_1+g_5 {\cal Z}_2+
\eta_4 \alpha_4+\eta_5 \alpha_5+\cdots \right)\,,
\label{LV3}
\ee
where $g_4, g_5, \eta_4, \eta_5$ are dimensionless constants.
We do not take into account the terms irrelevant to the discussion of 
linear cosmological perturbations on the flat FLRW 
background (such as ${\cal R}^3$).

Similarly, the Lagrangians corresponding to the $z=2$ and $z=1$ 
scalings are given, respectively, by
\ba
L_{{\cal V}_2}
&=& -\frac{1}{2}
\left( g_2 {\cal R}^2+g_3 {\cal Z}+
\eta_2 \alpha_2+\eta_3 \alpha_3+\cdots \right)\,,
\label{LV2}\\
L_{{\cal V}_1}
&=& \frac{M_{\rm pl}^2}{2} \left( {\cal R}
+\eta_1 \alpha_1 \right)\,,
\label{LV1}
\ea
where $g_2, g_3, \eta_1, \eta_2,\eta_3$ 
are dimensionless constants.
Summing up all the terms (\ref{LK})-(\ref{LV1}), 
the action of Ho\v{r}ava-Lifshitz gravity is characterized by 
 $S=\int d^4 x \sqrt{-g}\,L$ with the Lagrangian
\be
L =
\frac{M_{\rm pl}^2}{2} \left[ {\cal S}-\lambda K^2
+{\cal R}+\eta_1 \alpha_1
-M_{\rm pl}^{-2} \left( g_2 {\cal R}^2+g_3 {\cal Z}+
\eta_2 \alpha_2+\eta_3 \alpha_3\right)
-M_{\rm pl}^{-4}\left( g_4 {\cal Z}_1+g_5 {\cal Z}_2+
\eta_4 \alpha_4+\eta_5 \alpha_5 \right) \right]\,.
\label{Horavalag}
\ee
Since this Lagrangian depends on 
${\cal S}, K, {\cal R}, {\cal Z}, {\cal Z}_1, {\cal Z}_2, \alpha_i$ 
($i=1,2,\cdots, 5$), the theory belongs to the special case of (\ref{action}).
Note that the Lagrangian density ${\cal L}=\sqrt{-g}L=N \sqrt{h}\,L$ depends 
on the lapse $N$.

The original version of Ho\v{r}ava-Lifshitz gravity \cite{Horava}
corresponds to the case $N=N(t)$ and hence
$\alpha_i=0$. 
This scenario is plagued by pathological 
behavior associated with the instability of perturbations 
as well as the strong-coupling problem \cite{strong,strong2}.
These problems can be alleviated in the 
non-projectable extension of the theory \cite{Sergey,Blas}. 
In Sec.~\ref{apsec2} we shall discuss this issue after deriving 
the equations of linear cosmological perturbations.

\section{Equations of motion for the background and linear 
cosmological perturbations}
\label{persec} 

In this section we expand the action (\ref{action}) up to second order in 
perturbations on the flat FLRW background and derive the background 
and linear perturbation equations of motion.
The linear cosmological perturbations can be decomposed into 
scalar, vector, and tensor modes \cite{Bardeen,cosreview}, 
among which we focus on 
the dynamics of scalar perturbations in this paper. 
Let us consider the perturbed line element with four 
scalar variables $\delta N$, $\psi$, $\zeta$, 
and $E$, as
\be
ds^{2}=-(1+2\delta N) dt^{2}+2\nabla_i \psi dx^{i}dt+a^{2}(t) 
\left[ (1+2\zeta)\delta_{ij}+2\partial_{ij}E \right] 
dx^{i}dx^{j}\,,
\label{permet}
\ee
where $a(t)$ is the time-dependent scale factor, and 
$\partial_{ij} \equiv \nabla_i \nabla_j-\delta_{ij}\nabla^2/3$.
Under the infinitesimal transformation $t\rightarrow t+f(t,x^i)$ and 
$x^{i}\rightarrow x^{i}+\delta ^{ij}\nabla_{j}\xi (t,x^i)$, where 
$f$ and $\xi$ are scalar functions depending on $t$ and $x^i$, 
the perturbations $\delta N$ and $E$ transform as \cite{cosreview}
\ba
\delta N &\rightarrow& \delta N -\dot{f}\,,\label{gauge1}\\
E &\rightarrow& E-\xi\,.
\ea
The spatial gauge transformation is fixed by setting 
\be
E=0\,,
\label{E0}
\ee
whose condition is used throughout the paper.

Since in Horndeski and GLPV theories the
unitary gauge $\delta \phi=0$ is chosen, the dependence 
on a scalar field $\phi$ and its kinetic term $X$ does not explicitly 
appear in the action (\ref{action}). 
The transformation of the field perturbation is given by 
$\delta \phi \to \delta \phi-\dot{\phi}f$, so the temporal 
gauge transformation is fixed by choosing the unitary gauge.
We can deal with the action (\ref{action}) as if no field perturbations 
are present, but the propagating scalar degree of freedom manifests 
itself through the metric perturbations $\delta N$, $\psi$, and $\zeta$.
As we will see in Sec.~\ref{pereqsec}, the Hamiltonian and momentum constraints 
allow us to reduce the number of scalar variables further.

In the projectable version of Ho\v{r}ava-Lifshitz gravity where the lapse
$N$ is a function of $t$ alone, we have that $\delta N=0$. 
This is consistent with the foliation-preserving transformation 
$t \rightarrow t+f(t)$. 
In the non-projectable Ho\v{r}ava-Lifshitz gravity the lapse $N$ 
depends on the spatial coordinate $x^i$ as well as $t$, 
such that $\delta N=\delta N(x^i,t)$. 
Then the choice of the gauge $\delta N=0$ is inconsistent 
with the foliation-preserving transformation, as $f$ depends 
on $x^i$ from Eq.~(\ref{gauge1}).
In this case the temporal gauge transformation is not fixed,  
but it is possible to study the evolution of perturbations by 
appropriately constructing gauge-invariant variables 
(according to the line of Ref.~\cite{Urakawa}). 

In the following we expand the action (\ref{action}) up to second order 
in perturbations for the metric (\ref{permet}) 
with the gauge choice (\ref{E0}). 
On the flat FLRW background described by the line element 
$ds^2=-dt^2+a^2(t) \delta_{ij}dx^idx^j$, the extrinsic curvature
and the intrinsic curvature are given, respectively, by 
$\bar{K}_{ij}=H\bar{h}_{ij}$ and $\bar{{\cal R}}_{ij}=0$, 
where a bar represents background values and 
$H \equiv \dot{a}/a$ is the Hubble parameter.
Then, the scalar quantities 
appearing in the Lagrangian of (\ref{action}) read
\be
\bar{N}=1\,,\qquad
\bar{K}=3H\,,\qquad 
\bar{\s}=3H^{2}\,,
\qquad \bar{{\cal R}}=\bar{\Z}=\bar{\U}=0\,,\qquad
\bar{\Z}_1=\bar{\Z}_2=0\,,\qquad 
\bar{\alpha}_1=\bar{\alpha}_2=\cdots=\bar{\alpha}_5=0\,.
\label{backre}
\ee

We introduce the perturbed quantities
\be
\delta K_{\mu \nu}=K_{\mu \nu}-Hh_{\mu \nu}\,,\qquad
\delta K=K-3H\,,\qquad 
\delta \s=\s-3H^{2}=2H\delta K+\delta
K_{\nu}^{\mu} \delta K_{\mu}^{\nu}\,,
\label{delKS}
\ee
where the last equation arises from the first equation and 
the definition of $\s$. 
The scalar quantities $\cal{R}$ and $\Z$ associated 
with the intrinsic curvature appear only as perturbations. 
Up to quadratic order they can be expressed as
\be
\delta \mathcal{R}=\delta _{1}\mathcal{R}+\delta _{2}\mathcal{R}\,,
\qquad
\delta \Z=\delta \mathcal{R}_{\nu}^{\mu}\delta 
\mathcal{R}_{\mu}^{\nu}\,,
\ee
where $\delta _{1} \cal{R}$ and $\delta _{2} \cal{R}$ are first-order
and second-order perturbations in $\delta \cal{R}$, respectively. 
Clearly, $\delta \Z$ is a second-order quantity. 
{}From the first relation of Eq.~(\ref{delKS}), it follows that 
\be
\U=H{\cal R}+{\cal R}_{\nu}^{\mu}
\delta K_{\mu}^{\nu}\,,
\ee
where the first term on the r.h.s. involves the 
first-order contribution ($H \delta _{1}\mathcal{R}$) 
and the second-order contribution ($H\delta _{2}\mathcal{R}$), 
and the second term corresponds to
a second-order quantity. 
{}From the definition (\ref{Zdef}) and (\ref{aldef}) it is clear that the 
quantities $\Z_1$, $\Z_2$, $\alpha_i$ ($i=1,2,\cdots,5$) 
are second order in perturbations.

The above argument shows that the Lagrangian expanded up 
to second order is given by 
\begin{eqnarray}
L &=&\bar{L}+L_{,N}\delta N+L_{,K}\delta K+L_{,\s}
\delta \s+L_{,\mathcal{R}}\delta \mathcal{R}
+L_{,\Z}\delta \Z+L_{,\U}
\delta \U \nonumber \\
&&+\frac{1}{2}\left( \delta N\frac{\partial }{\partial N}
+\delta K\frac{\partial }{\partial K}+\delta \s
\frac{\partial }{\partial \s}+\delta \mathcal{R}
\frac{\partial }{\partial \mathcal{R}}+\delta \U
\frac{\partial }{\partial \U} \right)^2L
+\sum_{i=1}^2 L_{,\Z_i}\delta \Z_i
+\sum_{i=1}^5 L_{,\alpha_i}\delta \alpha_i+O(3)\,,
\label{lag}
\end{eqnarray}
where a comma represents a partial derivative, e.g., 
$L_{,N}=\partial L/\partial N$. 
Dividing the Lagrangian (\ref{lag}) into first-order and second-order 
contributions, we can obtain the equations of motion for 
the background and linear cosmological perturbations, respectively.

\subsection{Background equations of motion}

In order to derive the first-order Lagrangian, 
we first compute the combination $L_{,K}\delta K+L_{,\s}
\delta \s$ in Eq.~(\ref{lag}).
Making use of the second and third relations 
of Eq.~(\ref{delKS}) and defining the quantity
\be
{\cal F} \equiv L_{,K}+2HL_{,{\cal S}}\,,
\ee
it follows that 
\be
L_{,K}\delta K+L_{,\s}\delta \s ={\cal F}(K-3H)
+L_{,\s}\delta K_{\nu}^{\mu}\delta K_{\mu}^{\nu}\,.
\label{LKre}
\ee
Since $K=n_{~;\mu}^{\mu}$ from Eq.~(\ref{Kmunu}), the term 
${\cal F}K$ is partially integrated to give
\be
\int d^{4}x\sqrt{-g}\,\mathcal{F}K=-\int d^{4}x\sqrt{-g}\,
\mathcal{F}_{;\mu}n^{\mu}=-\int d^{4}x\sqrt{-g}\frac{\dot{\mathcal{F}}}{N}\,,
\label{delK}
\ee
up to a boundary term.
Expanding the term $N^{-1}=(1+\delta N)^{-1}$ 
up to second order, Eq.~(\ref{LKre}) reduces to
\be
L_{,K}\delta K+L_{,\s}\delta \s =
-\dot{{\cal F}}-3H\mathcal{F}+\dot{\mathcal{F}}\delta N
-\dot{\mathcal{F}}\delta N^{2}
+L_{,\s}\delta K_{\nu}^{\mu}\delta K_{\mu}^{\nu}+O(3)\,.  
\label{LKSre}
\ee

The first-order contribution to $L_{,\cal R}\delta {\cal R}$ 
of Eq.~(\ref{lag}) is given by  $L_{,\cal R}\delta_1 {\cal R}$, 
whereas $L_{,\Z}\delta \Z$ is second order. 
Employing the following relation \cite{Piazza}
\be
\int d^4 x \sqrt{-g}\,\lambda(t) \U=\int d^4 x\sqrt{-g} 
\left[\frac{\lambda(t)}{2} {\cal R}K+\frac{\dot{\lambda}(t)}
{2N} {\cal R} \right]\,,
\ee
which holds for a time-dependent function $\lambda(t)$ 
up to boundary terms, we obtain
\be
L_{,\U} \delta \U
= \frac{1}{2}\left( \dot{L_{,\U}}
+3HL_{,\U}\right) \delta _{1}\mathcal{R}
+\frac{1}{2}\left( \dot{L_{,\U}}
+3HL_{,\U}\right) \delta _{2}\mathcal{R}
+\frac{1}{2}\left( L_{,\U}\delta K-\dot{L_{,\U}}\delta N\right) \delta
_{1}\mathcal{R}+O(3) \,.
\label{LU}
\ee

In summary, the first-order action is given by $S=\int d^4 x\sqrt{-g}\,L$
with the Lagrangian
\be
L=\bar{L}-\dot{\mathcal{F}}-3H\mathcal{F}
+(\dot{\mathcal{F}}+L_{,N})\delta N
+\mathcal{E}\delta _{1}\mathcal{R}\,, 
\ee
where 
\be
\mathcal{E} \equiv L_{,\cal R}+\frac12 \dot{L_{,\U}}+
\frac32 H L_{,\U}\,.
\ee
We define the Lagrangian density as $\mathcal{L}=\sqrt{-g}L=N\sqrt{h}\,L$.
Then, the zeroth-order and first-order Lagrangian densities read
\begin{eqnarray}
\mathcal{L}_{0} &=& a^3 ( \bar{L}-\dot{\cal F}-3H {\cal F} )\,,
\label{lag0th} \\
\mathcal{L}_{1} &=&
a^{3} ( \bar{L}+L_{,N}-3H\mathcal{F} ) \delta N
+( \bar{L}-\dot{\mathcal{F}}-3H\mathcal{F} )
\delta \sqrt{h}+
a^3\mathcal{E} \delta_1 {\cal R}\,,\label{L1}
\end{eqnarray}
where the last term in Eq.~(\ref{L1}) is a total derivative.
Varying the first-order Lagrangian density (\ref{L1}) with respect to 
$\delta N$ and $\delta \sqrt{h}$, we obtain the background equations 
of motion
\begin{eqnarray}
&&\bar{L}+L_{,N}-3H\mathcal{F}=0\,,
\label{back1} \\
&&\bar{L}-\dot{\mathcal{F}}-3H\mathcal{F}=0\,.
\label{back2} 
\end{eqnarray}
The zero-th order Lagrangian density (\ref{lag0th}) vanishes 
on account of Eq.~(\ref{back2}). 
Combining Eq.~(\ref{back1}) with Eq.~(\ref{back2}), it follows that 
\be
L_{,N}+\dot{\mathcal{F}}=0\,.
\label{back3} 
\ee

Equation (\ref{back1}) corresponds to the Friedmann equation related to 
the Hubble parameter $H$, whereas Eq.~(\ref{back2}) is another 
independent equation associated with the time variation of $H$. 
In fact the Lagrangian of GR in the absence of matter is given by 
$L=(M_{\rm pl}^2/2)({\cal S}-K^2+{\cal R})$, in which case
$\bar{L}=-3M_{\rm pl}^2H^2$, $L_N=0$, and ${\cal F}=-2M_{\rm pl}^2 H$. 
Note that, for the theories with $N=N(t)$, one cannot 
take the variation with respect to $\delta N$. 
Indeed, this happens for the projectable version of Ho\v{r}ava Lifshitz gravity
(see Sec.\,\ref{apsec2}).
In the presence of a matter fluid with energy density $\rho_m$ and pressure $P_m$, 
the r.h.s. of Eqs.~(\ref{back1}), (\ref{back2}), and 
(\ref{back3}) are modified as $\rho_m$, $-P_m$, and $\rho_m+P_m$, 
respectively (see Sec.\,\ref{DEsec}).

\subsection{Perturbation equations of motion}
\label{pereqsec}

Now we explicitly compute the Lagrangian (\ref{lag}) to 
derive linear perturbation equations of motion. 
On using the relations (\ref{LKSre}) and (\ref{LU}) as well, the resulting 
Lagrangian can be expressed as
\begin{eqnarray}
L &=&\bar{L}-\dot{\mathcal{F}}-3H\mathcal{F}
+(\dot{\mathcal{F}}+L_{,N})\delta
N+\mathcal{E}\delta _{1}\mathcal{R} \nonumber \\
&&+\left( \frac{1}{2}L_{,NN}-\dot{\mathcal{F}}\right) \delta N^{2}
+\frac{1}{2}\A \delta K^{2}+\B \delta K\delta N
+\C \delta K\delta _{1}\mathcal{R}
+{\cal D} \delta N\delta _{1}\mathcal{R} 
+\mathcal{E}\delta _{2}\mathcal{R}+\frac{1}{2} 
\G \delta _{1}\mathcal{R}^{2} \nonumber \\
&&
+L_{,\s}\delta K_{\nu}^{\mu}\delta K_{\mu}^{\nu} 
+L_{,\Z}\delta \mathcal{R}_{\nu}^{\mu} \delta 
\mathcal{R}_{\mu}^{\nu}+\sum_{i=1}^2 L_{,\Z_i}\delta \Z_i
+\sum_{i=1}^5 L_{,\alpha_i}\delta \alpha_i+O(3)\,, 
 \label{lagex}
\end{eqnarray}
where 
\begin{eqnarray}
\A &=&L_{,KK}+4HL_{,\s K} +4H^{2}L_{,\s \s}\,,\\
\B &=&L_{,KN}+2HL_{,\s N}\,, \\
\C &=&L_{,K\mathcal{R}}+2HL_{,\s \mathcal{R}}
+\frac{1}{2} L_{,\U}+HL_{,K \U}+2H^{2}L_{,\s \U}\,, \\
{\cal D} &=&L_{,N\mathcal{R}}-\frac{1}{2}\dot{L_{,\U}}
+HL_{,N \U}\,, \\
\G &=&L_{,\mathcal{R}\mathcal{R}}
+2HL_{,\mathcal{R}\U}+H^{2}L_{,\U \U}\,.
\end{eqnarray}
Denoting the first-order and second-order Lagrangians 
of Eq.~(\ref{lagex}) as $L_1$ and $L_2$, respectively, 
the second-order Lagrangian density can be 
evaluated as ${\cal L}_2=a^3\delta N L_1+\delta \sqrt{h}\,L_1+a^3L_2$, i.e., 
\begin{eqnarray}
\hspace{-0.5cm}
\mathcal{L}_{2} &=&\delta \sqrt{h}
\left[ (\dot{\mathcal{F}}+L_{,N})\delta N+ 
\mathcal{E}\delta _{1}\mathcal{R}\right]  \nonumber \\
\hspace{-0.5cm}
&&+a^{3}\biggl[\left( L_{,N}+\frac{1}{2}L_{,NN}\right) \delta N^{2}+\mathcal{E}
\delta _{2}\mathcal{R}+\frac{1}{2} \A \delta K^{2}+\B \delta
K\delta N+\C \delta K\delta _{1}\mathcal{R}
+({\cal D}+\mathcal{E})\delta N\delta _{1}\mathcal{R}
+\frac{1}{2} \G \delta _{1}\mathcal{R}^{2} \nonumber \\
& &~~~~~~~
+L_{,\s}\delta K_{\nu}^{\mu }\delta K_{\mu }^{\nu} 
+L_{,\Z}\delta \mathcal{R}_{\nu}^{\mu} \delta 
\mathcal{R}_{\mu}^{\nu}+\sum_{i=1}^2 L_{,\Z_i}\delta \Z_i
+\sum_{i=1}^5 L_{,\alpha_i}\delta \alpha_i \biggr]\,. 
\label{L2den}
\end{eqnarray}
Here we expanded $\sqrt{-g}$ up to first order, since the 
second-order contribution is multiplied by the zeroth-order Lagrangian 
(\ref{lag0th}) and it vanishes identically due to Eq.~(\ref{back2}). 

The next step is to express the perturbed quantities such as 
$\delta K$ and $\delta_1 {\cal R}$ in terms of metric perturbations 
$\delta N$, $\psi$, and $\zeta$.
We recall that the extrinsic curvature is given by Eq.~(\ref{Kijm2}).
Since $h_{ij}=a^2(t) (1+2\zeta) \delta_{ij}$ for the gauge choice 
(\ref{E0}), the first-order extrinsic curvature can be expressed as
\be
\delta K_{j}^{i}=( \dot{\zeta}-H\delta N ) \delta _{j}^{i} -\frac{1
}{2a^{2}}\delta ^{ik}(\partial _{k}N_{j}+\partial _{j}N_{k})\,.
\label{Kij}
\ee
Here the three-dimensional derivatives like $\nabla_i N_j$ 
have been replaced by partial derivatives like $\partial_i N_j$, 
as the Christoffel symbols $\Gamma _{ij}^{k}$ are first-order 
perturbations. 
Using the property $N_i= \partial_i \psi$ and taking 
the trace of Eq.~(\ref{Kij}), we obtain
\be
\delta K=3( \dot{\zeta}-H\delta N ) 
-\Delta \psi \,, 
\label{Ktrace}
\ee
where 
\be
\Delta=\nabla_i \nabla^i=\frac{1}{a^2(t)}\delta^{ij} 
\partial_i \partial_j \equiv \frac{1}{a^2(t)} \partial^2\,.
\ee
The operator $\Delta$ involves the scale factor $a(t)$, 
so we need to be careful when we take time derivatives 
of the quantities like $\Delta \psi$.

In Eq.~(\ref{L2den}) the perturbation $\delta \sqrt{h}$ is 
equivalent to $3a^3 \zeta$. 
The perturbations of the intrinsic curvature are 
given by 
\be
\delta \mathcal{R}_{ij}=-\left( \delta
_{ij}\partial ^{2}\zeta +\partial _{i}\partial _{j}\zeta \right) \,,
\qquad \delta _{1}\mathcal{R}=-4a^{-2}\partial ^{2}\zeta \,,\qquad \delta _{2} 
\mathcal{R}=-2a^{-2}\left[ (\partial \zeta )^{2}-4\zeta \partial ^{2}\zeta 
\right]\,,
\label{delR}
\ee
where $(\partial \zeta )^{2} \equiv \delta^{ij} (\partial_{i}\zeta )(\partial _{j}\zeta)$.
On using these relations, the term $L_{,\Z_1}\delta \Z_1$ can be 
evaluated as
\be
L_{,\Z_1}\delta \Z_1=L_{,\Z_1} (\nabla^i \delta_1 {\cal R})
(\nabla_i \delta_1 {\cal R})
=16L_{,\Z_1} (\nabla^i \Delta \zeta) (\nabla_i \Delta \zeta)
=16L_{,\Z_1} (\Delta \partial^i \zeta) (\Delta \partial_i\zeta)\,,
\label{LZ1}
\ee
which is valid up to second order in perturbations.
Similarly, we have 
\be
L_{,\Z_2}\delta \Z_2=L_{,\Z_2} 
(h_{jk}\Delta \partial_i \zeta+ \partial_i  \partial_j
\partial_k \zeta)(h^{jk}\Delta \partial^i \zeta+ 
\partial^i  \partial^j \partial^k \zeta)
=6L_{,\Z_2} (\Delta \partial^i \zeta) (\Delta \partial_i\zeta)\,, 
\ee
up to a boundary term. 
The perturbed quantities associated 
with the acceleration read
\ba
L_{,\alpha_1}\delta \alpha_1
&=& L_{,\alpha_1} (\partial_i \delta N)(\partial^i \delta N)\,,\\
L_{,\alpha_2}\delta \alpha_2
&=& L_{,\alpha_2} (\partial_i \delta N) \Delta (\partial^i \delta N)\,,\\
L_{,\alpha_3}\delta \alpha_3
&=& 4L_{,\alpha_3} (\partial_i \delta N) \Delta (\partial^i \zeta)\,,\\
L_{,\alpha_4}\delta \alpha_4
&=& L_{,\alpha_4} \Delta (\partial_i \delta N) \Delta (\partial^i \delta N)\,,\\
L_{,\alpha_5}\delta \alpha_5
&=& 4L_{,\alpha_5} \Delta (\partial_i \delta N) \Delta (\partial^i \zeta)\,.
\label{Lal}
\ea

Substituting the relations (\ref{Kij}), (\ref{Ktrace}), (\ref{delR}), and 
(\ref{LZ1})-(\ref{Lal}) into Eq.~(\ref{L2den}) and using the background 
equation (\ref{back3}), it follows that  
\begin{eqnarray}
\hspace{-0.3cm}
\mathcal{L}_2 &=& a^3 \biggl\{ \frac12 \left[ 
2L_{,N}+L_{,NN}-6H{\cal W}
-3H^{2}(3 \A+2L_{,\s}) \right] \delta N^2 
+\left[ {\cal W} ( 3\dot{\zeta}-\Delta \psi ) 
+4(3H \C-{\cal D}-\mathcal{E}) \Delta \zeta \right] \delta N 
\nonumber \\
\hspace{-0.3cm}
& &~~~-(3 \A+2L_{,\s}) \dot{\zeta} \Delta \psi 
-12\C \dot{\zeta} \Delta \zeta
+\left( \frac92 {\cal A}+3L_{,\s} \right) \dot{\zeta}^2 +2\mathcal{E} 
\frac{(\partial \zeta)^2}{a^2}  \nonumber \\
\hspace{-0.3cm}
& &~~~+\frac12 ( \A+2L_{,\s} ) (\Delta \psi)^2
+4\C (\Delta \psi)(\Delta \zeta) 
+2(4\G+3L_{,\Z}) (\Delta \zeta)^2 
\nonumber \\
\hspace{-0.3cm}
& &~~~+2 (8L_{,\Z_1}+3L_{,\Z_2})
(\Delta \partial^i \zeta) (\Delta \partial_i\zeta)
+L_{,\alpha_1} (\partial_i \delta N)(\partial^i \delta N)
+L_{,\alpha_2} (\partial_i \delta N) \Delta (\partial^i \delta N)
\nonumber \\
\hspace{-0.3cm}
& &~~~
+4L_{,\alpha_3} (\partial_i \delta N) \Delta (\partial^i \zeta)
+L_{,\alpha_4} \Delta (\partial_i \delta N) \Delta (\partial^i \delta N)
+4L_{,\alpha_5} \Delta (\partial_i \delta N) \Delta (\partial^i \zeta)
\biggr\}\,,
\label{L2exp}
\end{eqnarray}
where 
\be
{\cal W} \equiv \B-3\A H-2L_{,\s}H\,.
\ee 

The Lagrangian density (\ref{L2exp}) involves spatial derivatives 
higher than second order, so its variation 
with respect to $\delta N$ corresponds to 
the equation of motion
\be
\frac{\partial {\cal L}_2}{\partial (\delta N)}
-\partial_i \left( \frac{\partial {\cal L}_2}{\partial (\partial_i \delta N)} 
\right)+\partial_i \partial_j  
\left( \frac{\partial {\cal L}_2}{\partial (\partial_i \partial_j \delta N)} \right)
-\partial_i \partial_j  \partial_k
\left( \frac{\partial {\cal L}_2}{\partial (\partial_i \partial_j \partial_k \delta N)} \right)
+\cdots=0\,.
\label{Hamio}
\ee
For example, the term $a^3L_{,\alpha_2} (\partial_i \delta N) \Delta (\partial^i \delta N)$ 
is equivalent to $-a^3L_{,\alpha_2} (\partial_i \partial_j \delta N)
(\partial^i \partial^j \delta N)=-a^3L_{,\alpha_2}(\partial_i \partial_j \delta N)^2/a^4$ 
up to a boundary term, 
so that the third term on the l.h.s. of (\ref{Hamio}) gives rise to the contribution 
$-2a^3L_{,\alpha_2} \Delta^2 \delta N$. 
Taking the variations of other terms as well, Eq.~(\ref{Hamio}) 
leads to the following Hamiltonian constraint:
\ba
& &
\left[ 2L_{,N}+L_{,NN}-6H{\cal W}
-3H^{2}(3 \A+2L_{,\s}) \right] \delta N
+ {\cal W} ( 3\dot{\zeta}-\Delta \psi ) 
+4(3H \C-{\cal D}-\mathcal{E}) \Delta \zeta \nonumber \\
& &
-2L_{,\alpha_1} \Delta \delta N
-2L_{,\alpha_2} \Delta^2 \delta N
-4L_{,\alpha_3} \Delta^2 \zeta
-2L_{,\alpha_4} \Delta^3 \delta N
-4L_{,\alpha_5} \Delta^3 \zeta=0\,.
\label{Hami}
\ea

Varying the Lagrangian (\ref{L2exp}) with respect to $\psi$ 
gives the following momentum constraint:
\be
{\cal W}\delta N+(3{\cal A}+2L_{,\s}) \dot{\zeta}
-4{\cal C} \Delta \zeta-({\cal A}+2L_{,\s})\Delta \psi=0\,.
\label{momen}
\ee
Finally, the variation of (\ref{L2exp}) in terms of $\zeta$ 
leads to
\ba
& &
\dot{{\cal Y}}+3H {\cal Y}-4(3H \C-{\cal D}-\mathcal{E}) \Delta \delta N
+4{\cal E} \Delta \zeta +12{\cal C} \Delta \dot{\zeta} 
\nonumber \\
& &
-4{\cal C} \Delta^2 \psi-4 (4{\cal G}+3L_{,\Z}) \Delta^2 \zeta
+4(8L_{,\Z_1}+3L_{,\Z_2})\Delta^3 \zeta
+4L_{,\alpha_3} \Delta^2 \delta N+4L_{,\alpha_5} \Delta^3 \delta N
=0\,,
\label{zeta}
\ea
where 
\be
{\cal Y} \equiv 3\left[ {\cal W} \delta N +
(3{\cal A}+2L_{,\s}) \dot{\zeta}
-4{\cal C} \Delta \zeta \right]
-(3{\cal A}+2L_{,\s})\Delta \psi\,.
\label{Ydef}
\ee
Using the momentum constraint (\ref{momen}), the quantity (\ref{Ydef}) 
reduces to 
\be
{\cal Y} =4L_{,\s} \Delta \psi=4L_{,\s}\frac{\partial^2 \psi}{a^2}\,,
\label{Y2}
\ee
and hence 
$\dot{{\cal Y}}+3H {\cal Y}=\Delta [4(L_{,\s}\dot{\psi}+\dot{L}_{,\s}\psi+HL_{,\s}\psi)]$.
Then, Eq.~(\ref{zeta}) can be expressed in the following form
\ba
& &
\Delta [L_{,\s}\dot{\psi}+\dot{L}_{,\s}\psi+HL_{,\s}\psi
+{\cal E} \zeta+3{\cal C}\dot{\zeta}
-(3H \C-{\cal D}-\mathcal{E})\delta N \nonumber \\
& &~~~
-{\cal C}\Delta \psi-(4{\cal G}+3L_{,\Z}) \Delta \zeta
+(8L_{,\Z_1}+3L_{,\Z_2})\Delta^2 \zeta
+L_{,\alpha_3} \Delta \delta N+L_{,\alpha_5} \Delta^2 
\delta N]=0\,.
\label{zeta2}
\ea
The dynamics of linear cosmological perturbations is known 
by solving Eqs.~(\ref{Hami}), (\ref{momen}), and (\ref{zeta}) [or (\ref{zeta2})]
together with the background equations of motion
(\ref{back1}) and (\ref{back2}).

\subsection{Second-order linear perturbations}
\label{secondsec}

The linear perturbation equations (\ref{Hami})-(\ref{zeta}) 
involve time derivatives up to second order, but there are 
spatial derivatives higher than two. 
Let us consider second-order theory in which spatial derivatives and the 
combination of time and spatial derivatives remain of second order.

The higher-order spatial derivative terms on the second line 
of Eq.~(\ref{zeta}) are absent under the conditions
${\cal C}=0$, $4{\cal G}+3L_{,\Z}=0$, $8L_{,\Z_1}+3L_{,\Z_2}=0$, 
$L_{,\alpha_3}=0$, and $L_{,\alpha_5} =0$. 
In addition we also require $L_{,\alpha_2}=0$ and $L_{,\alpha_4} =0$ 
for the absence of higher-order derivatives in Eq.~(\ref{Hami}).
Provided that ${\cal W} \neq 0$, the perturbation $\delta N$ 
is related with the combination $({\cal A}+2L_{,\s})\Delta \psi$. 
Then, the third term on the l.h.s. of Eq.~(\ref{zeta}) induces the 
fourth-order spatial derivative $\Delta^2 \psi$, which can be 
eliminated under the condition ${\cal A}+2L_{,\s}=0$.
As we will see in Sec.~\ref{apsec1}, this condition is satisfied 
in Horndeski theory, but not in Ho\v{r}ava-Lifshitz gravity. 
This makes sense as Ho\v{r}ava-Lifshitz gravity involves 
spatial derivatives higher than two by construction.

{}From Eq.~(\ref{momen}) the perturbation $\delta N$ also 
depends upon the term $(3{\cal A}+2L_{,\s}) \dot{\zeta}$ for 
${\cal W} \neq 0$. The third term on the l.h.s. of Eq.~(\ref{zeta}) 
gives rise to the combination of time and spatial derivatives 
$\Delta \dot{\zeta}$ higher than two, 
but this exactly cancels another term
(as we will see below). 
On the other hand, the term $-2L_{,\alpha_1} \Delta \delta N$ 
in Eq.~(\ref{Hami}) does not vanish unless the condition 
$L_{,\alpha_1}=0$ is satisfied.

In summary, the spatial derivatives and the combination of 
time and spatial derivatives higher than second order 
are absent for linear perturbations under the conditions
\be
{\cal C}=0\,,\qquad 
4{\cal G}+3L_{,\Z}=0\,,\qquad
{\cal A}+2L_{,\s}=0\,, \qquad
8L_{,\Z_1}+3L_{,\Z_2}=0\,,\qquad
L_{,\alpha_1}=L_{,\alpha_2}=\cdots=L_{,\alpha_5}=0\,.
\label{nospa}
\ee
We note that, even if these conditions are satisfied, 
higher-order derivative terms may be present 
for {\it non-linear} perturbations \cite{Piazza}.

Under the conditions (\ref{nospa}), Eqs.~(\ref{Hami}), 
(\ref{momen}), and (\ref{zeta2}) reduce, respectively, to 
\ba
& &
\left( 2L_{,N}+L_{,NN}-6H{\cal W}+12H^2 L_{,\s} \right) \delta N
+ {\cal W} ( 3\dot{\zeta}-\Delta \psi ) 
-4({\cal D}+\mathcal{E}) \Delta \zeta=0\,,
\label{coper1} \\
& & 
{\cal W} \delta N-4L_{,\s} \dot{\zeta}=0\,,
\label{coper2} \\
& &
\frac{1}{a^3} \frac{d}{dt} (a^3 L_{,\s} \Delta \psi)
+{\cal E} \Delta \zeta+({\cal D}+{\cal E}) \Delta \delta N=0\,.
\label{coper3}
\ea
Provided that ${\cal W}=L_{,KN}+2HL_{,\s N}+4HL_{,\s} \neq 0$, we have 
$\delta N=4L_{,\s} \dot{\zeta}/{\cal W}$ from Eq.~(\ref{coper2}). 
Substituting this relation into Eq.~(\ref{coper1}), we obtain 
\be
\Delta \psi=\frac{Q_s}{2L_{,\s}} \dot{\zeta}
-\frac{4({\cal D}+{\cal E})}{{\cal W}} \Delta \zeta\,,
\label{delpsi}
\ee
where 
\ba
Q_s &\equiv& \frac{2L_{,\s}}{{\cal W}^2} 
\left[ 3{\cal W}^2+4L_{,\s} 
(2L_{,N}+L_{,NN}-6H{\cal W}+12H^2L_{,\s}) \right]\,.
\label{Qsdef}
\ea
Expressing the terms $\Delta \delta N$ and $\Delta \psi$
of Eq.~(\ref{coper3}) in terms of $\zeta$ and its derivatives, 
we find that the two terms $\Delta \dot{\zeta}$ cancels out 
and that the resulting equation of motion is given by 
\be
\frac{d}{dt} \left( a^3 Q_s \dot{\zeta} \right)-a Q_s c_s^2
\partial^2 \zeta=0\,,
\label{pereq}
\ee
where
\be
c_s^2 \equiv \frac{2}{Q_s} 
\left( \dot{{\cal M}}+H {\cal M}-\mathcal{E} \right)\,,
\label{cs}
\ee
with 
\be
{\cal M} \equiv \frac{4L_{,\s} ({\cal D}+{\cal E})}{{\cal W}}
=\frac{4L_{,\s}}{{\cal W}} \left(
L_{,\cal R}+L_{,N{\cal R}}+HL_{,N \U}
+\frac32 HL_{,\U} \right)\,. 
\label{Mdef}
\ee

Substituting the relations (\ref{coper2}) and (\ref{delpsi}) 
into Eq.~(\ref{L2exp}) for ${\cal W} \neq 0$, 
the second-order Lagrangian density reads
\be
\mathcal{L}_2=a^3 Q_s \left[ \dot{\zeta}^2-\frac{c_s^2}{a^2}
(\partial \zeta)^2 \right]\,.
\label{perlag}
\ee
In fact, this gives rise to the equation of motion (\ref{pereq}) 
for the curvature perturbation $\zeta$. 
The above results show that, under the conditions (\ref{nospa}), 
there is one scalar propagating degree of freedom
with second-order equations of motion.
The scalar ghost is absent for 
\be
Q_s>0\,.
\label{consi1}
\ee
In order to avoid small-scale instabilities associated with the 
Laplacian term $\partial^2 \zeta$ in Eq.~(\ref{pereq}),  
we also require
\be
c_s^2>0\,.
\label{consi2}
\ee
The two conditions (\ref{consi1}) and (\ref{consi2}) need to be satisfied 
for the consistency of second-order gravitational theory.

\section{Application to Horndeski and GLPV theories}
\label{apsec1} 

Let us first apply the results in the previous section to 
Horndeski and GLPV theories.
We recall that the spatial gauge transformation is fixed 
by choosing the gauge $E=0$.
We also choose the unitary gauge $\delta \phi=0$ 
to fix the temporal transformation. 
In this case, the Horndeski Lagrangian (\ref{Lagsum}) with 
(\ref{eachlag2})-(\ref{L5lag}) reduces to the form (\ref{LH2}) 
with the coefficients (\ref{AB}). 
Horndeski theory has the two restrictions (\ref{ABcon}) 
among the coefficients, while GLPV theory is 
described by the Lagrangian (\ref{LH2}) with 6 arbitrary functions 
$A_2,A_3,A_4,A_5$ and $B_4,B_5$.

\subsection{Background equations of motion}
\label{basec}

On using the properties (\ref{backre}) and the relations 
$\phi=\phi(t)$ and $X=-\dot{\phi}^2(t)/N^2$
in unitary gauge, the background equations 
of motion (\ref{back1}) and (\ref{back2}) 
in GLPV theory read 
\ba
\hspace{-0.5cm}
& & A_2-6H^2 A_4-12H^3 A_5+2\dot{\phi}^2 
\left( A_{2,X}+3H A_{3,X}+6H^2 A_{4,X}+6H^3 A_{5,X} \right)=0\,,
\label{back1d} \\
\hspace{-0.5cm}
& & A_2-6H^2 A_4-12H^3 A_5-\dot{A}_3-4\dot{H}A_4
-4H\dot{A}_4-12H \dot{H}A_5-6H^2 \dot{A}_5=0\,,
\label{back2d} 
\ea
respectively. 
In Horndeski theory the equations of motion can be derived by 
substituting the functions (\ref{AB}) into Eqs.~(\ref{back1d}) 
and (\ref{back2d}).
In the presence of a perfect fluid with energy density 
$\rho_m$ and pressure $P_m$, we need to add the 
terms $\rho_m$ and $-P_m$ on the r.h.s. of 
Eqs.~(\ref{back1d}) and (\ref{back2d}) respectively 
(see Sec.\ \ref{DEsec}).

Since Eqs.~(\ref{back1d}) and (\ref{back2d}) 
do not contain the functions $B_4$ and $B_5$, the theories 
with same values of $A_2, A_3, A_4, A_5$ but with different values of 
$B_4, B_5$ cannot be distinguished from each other at the background level.
To be concrete, let us consider the covariant Galileon 
theory \cite{Galileon} described by the following functions 
in the Horndeski Lagrangians (\ref{eachlag2})-(\ref{L5lag}):
\be
G_2=\frac{c_2}{2}X\,,\qquad
G_3=\frac{c_3}{2M^3}X\,,\qquad
G_4=\frac{M_{\rm pl}^2}{2}-\frac{c_4}{4M^6}X^2\,,\qquad
G_5=\frac{3c_5}{4M^9}X^2\,,
\label{Gali1}
\ee
where $c_i$ ($i=2,3,4,5$) are dimensionless constants, and 
$M$ is a constant having a dimension of mass.
The auxiliary functions $F_3$ and $F_5$ appearing in 
Eqs.~(\ref{F3}) and (\ref{F5}) can be chosen as 
$F_3=c_3X/(6M^3)$ and $F_5=3c_5X^2/(5M^9)$, respectively.
{}From Eq.~(\ref{AB}) the functions appearing in the 
Lagrangian (\ref{LH2}) for the covariant Galileon are given by 
\ba
& & 
A_2=\frac{c_2}{2}X\,,\qquad 
A_3=\frac{c_3}{3M^3} (-X)^{3/2}\,,\qquad
A_4=-\frac{M_{\rm pl}^2}{2}-\frac{3c_4}{4M^6}X^2\,,\qquad
A_5=\frac{c_5}{2M^9} (-X)^{5/2}\,,\nonumber \\
& &
B_4=\frac{M_{\rm pl}^2}{2}-\frac{c_4}{4M^6}X^2\,,\qquad
B_5=-\frac{3c_5}{5M^9}(-X)^{5/2}\,.
\label{cova1}
\ea
Substituting the four functions $A_2,A_3,A_4,A_5$ into 
Eqs.~(\ref{back1d}) and (\ref{back2d}), the resulting 
background equations of motion match with those 
derived in Refs.~\cite{GS,Gacosmo} by direct variation 
of the Lagrangians (\ref{eachlag2})-(\ref{L5lag}) 
with the functions (\ref{Gali1}).

The covariant Galileon discussed above corresponds to 
the second-order theory in curved space-time 
(i.e., it belongs to a class of Horndeski theory). 
In the limit of Minkowski space-time, the equations of 
motion for the covariant Galileon are invariant under the 
Galilean shift $\partial_{\mu} \phi \to \partial_{\mu}\phi+b_{\mu}$. 
In fact, the original Galileon theory was 
constructed by Nicolis {\it et al.} \cite{Nicolis} to satisfy this 
Galilean symmetry in Minkowski space-time.
If we replace partial derivatives of the Minkowski Galileon 
with covariant derivatives (``covariantized Galileons''), this generally 
gives rise to derivatives higher than second order \cite{Galileon}. 
In fact, the covariant Galileon was elaborated to keep the 
equations of motion up to second order
by adding a counter term to the covariantized 
version of the original Galileon theory.

Although the covariantized Galileon theory contains derivatives 
higher than two in general space-time, this is not 
the case for the flat FLRW background. 
In the presence of the Einstein-Hilbert term 
$(M_{\rm pl}^2/2)R$ the functions $A_2,A_3,A_4,A_5$ 
for the covariantized Galileon are the same as 
those given in Eq.~(\ref{cova1}), but the 
functions $B_4$ and $B_5$ are different:
\be
B_4=\frac{M_{\rm pl}^2}{2}\,,\qquad
B_5=0\,.
\ee
Compared to Eq.~(\ref{cova1}) the gravitational 
counter terms $-c_4X^2/(4M^6)$ and 
$-3c_5(-X)^{5/2}/(5M^9)$ are absent, but 
the same second-order background equations 
as those for the covariant Galileon 
follow from Eqs.~(\ref{back1d}) and (\ref{back2d}). 
This second-order property also holds for linear 
cosmological perturbations on the flat FLRW background
(as we will see in Sec.\ \ref{DEsec}), 
but the difference between covariant and covariantized Galileons 
arises at the level of perturbations \cite{Kase}.

\subsection{Cosmological perturbations}

Let us proceed to the discussion of linear cosmological perturbations.
One can easily show that the Lagrangian (\ref{LH2}) of 
GLPV theory satisfies all the conditions (\ref{nospa}), so 
the perturbation equations of motion on the flat FLRW background 
remain of second order. Then, the discussion given in 
Sec.~\ref{secondsec} can be applied to GLPV theory.  
We recall that Eq.~(\ref{pereq}) is valid for 
\be
{\cal W}=A_{3,N}+4HA_{4,N}+6H^2A_{5,N}
-4H A_4-12H^2A_5 \neq 0\,.
\label{calW}
\ee
In Horndeski theory the functions $A_3,A_4,A_5$ are
given by Eq.~(\ref{AB}) with the correspondence 
$X=-\dot{\phi}^2(t)/N^2$ in unitary gauge, so
the condition (\ref{calW}) translates to
\ba
{\cal W}^{{\rm (H)}}
&=&
4HG_{4}+2\dot{\phi}XG_{3,X}-16H(XG_{4,X}+X^{2}G_{4,XX}) 
+2\dot{\phi}(G_{4,\phi}+2XG_{4,\phi X})\nonumber \\
&&-2H^{2}\dot{\phi}(5XG_{5,X}+2X^{2}G_{5,XX}) 
-2HX(3G_{5,\phi}+2XG_{5,\phi X})
\neq 0\,,
\label{calW2}
\ea
where the index ``(H)'' represents the values in Horndeski theory.
In GR with $G_4=M_{\rm pl}^2/2$ and 
$G_3=G_5=0$ we have ${\cal W}^{{\rm (H)}}
=2M_{\rm pl}^2 H$, so ${\cal W}^{{\rm (H)}}$ 
does not vanish on the cosmological background.

The quantity $Q_s$ defined in Eq.~(\ref{Qsdef}) 
can be expressed as
\be
Q_s=\frac{2L_{,\s}}{3{\cal W}^2} 
\left( 9{\cal W}^2+8L_{,\s} w \right)\,,
\label{Qsho}
\ee
where 
\ba
L_{,\s} &=& -A_4-3HA_5\,, \label{Lsdef} \\
w &\equiv& 3L_{,N}+3L_{,NN}/2-9H{\cal W}+18H^2 L_{,\s}  \nonumber \\
&=& 18H^2 (A_4+3HA_5)+3(A_{2,N}-6H^2 A_{4,N}-12H^3 A_{5,N}) 
\nonumber \\
& &+\frac32 (A_{2,NN}+3HA_{3,NN}+6H^2A_{4,NN}+6H^3A_{5,NN})\,.
\label{wdef}
\ea
Notice that the quantities ${\cal W}$, $L_{,\s}$, and $w$ appearing in 
$Q_s$ do not depend 
on the functions $B_4$ and $B_5$.
Hence the no-ghost conditions for the theories with same 
values of $A_2,A_3,A_4,A_5$ but with different values of 
$B_4,B_5$ (like covariant/covariantized Galileons) are 
equivalent to each other.
In Horndeski theory, Eqs.~(\ref{Lsdef}) and (\ref{wdef}) reduce to
\ba
L_{,\s}^{{\rm (H)}} &=& 
G_{4}-2XG_{4,X}-H\dot{\phi}XG_{5,X} 
-\frac12 XG_{5,\phi}\,, 
\label{Ls2}
\\
w^{{\rm (H)}} &=&-18H^{2}G_{4}+3(XG_{2,X}+2X^{2}G_{2,XX})
-18H\dot{\phi}(2XG_{3,X}+X^{2}G_{3,XX})
-3X(G_{3,\phi}+XG_{3,\phi X}) \nonumber \\
&& 
+18H^{2}(7XG_{4,X}+16X^{2}G_{4,XX}+4X^{3}G_{4,XXX})
-18H\dot{\phi}(G_{4,\phi}+5XG_{4,\phi X}+2X^{2}G_{4,\phi XX})  
 \nonumber \\
&& 
+6H^{3}\dot{\phi}(15XG_{5,X}+13X^{2}G_{5,XX}+2X^{3}G_{5,XXX})
+9H^{2}X(6G_{5,\phi}+9XG_{5,\phi X}+2X^{2}G_{5,\phi XX})\,.
\ea

The functions ${\cal E}$ and ${\cal D}+{\cal E}$ appearing in the 
scalar propagation speed square (\ref{cs}) are 
given, respectively, by 
\ba
{\cal E} &=& B_4+\frac12 \dot{B}_5\,,\\
{\cal D}+{\cal E} &=&
B_4+B_{4,N}-\frac12 H B_{5,N}\,. 
\ea
Hence the theories with different values of $B_4,B_5$ give rise to 
different scalar propagation speeds. 
In Horndeski theory there exist the particular relations (\ref{ABcon}) 
between the coefficients $A_4,A_5,B_4,B_5$.
In this case we have
\be
L_{,\s}^{{\rm (H)}}={\cal D}^{{\rm (H)}}+{\cal E}^{{\rm (H)}}\,,\qquad 
{\cal M}^{{\rm (H)}}=\frac{4{L_{,\s}^{{\rm (H)}}}^2}{{\cal W}^{{\rm (H)}}}\,.
\label{Mre}
\ee
Under the no-ghost condition $Q_s>0$, the condition 
(\ref{consi2}) for the absence of Laplacian instabilities 
in Horndeski theory translates to 
\be
\frac{1}{a}\frac{d}{dt} \left( a {\cal M}^{{\rm (H)}} \right)
-{\cal E}^{\rm (H)}>0\,,
\ee
where 
\be
{\cal E}^{\rm (H)}=G_{4}+\frac{1}{2}XG_{5,\phi}-XG_{5,X}\ddot{\phi}\,.
\label{Eho}
\ee

If we consider tensor perturbations $\gamma_{ij}$ in addition 
to scalar perturbations, this gives rise to additional conditions 
for the absence of ghosts and Laplacian instabilities.
The three-dimensional metric involving traceless and divergence-free 
tensor modes (satisfying $\gamma _{ii}=\partial _{i}\gamma _{ij}=0$)
can be written as \cite{Maldacena}
\be
h_{ij}=a^{2}(t)(1+2\zeta)\hat{h}_{ij}\,,\qquad 
\hat{h}_{ij}=\delta_{ij}+\gamma _{ij}
+\frac{1}{2}\gamma _{ik}\gamma _{kj}\,,\qquad \mathrm{det}
\,\hat{h}=1\,,  \label{gra}
\ee
where the term $\gamma _{ik}\gamma _{kj}/2$ 
has been introduced for simplifying the calculations.
At linear order, the tensor modes decouple from the scalar 
modes and they satisfy the relations 
$\delta K=0$, $\delta K^{i}_{j}=\delta^{ik} \dot{\gamma}_{kj}/2$, 
$\delta_{1}\mathcal{R}=0$, and 
$\delta _{2}\mathcal{R}=-(\partial _{k}\gamma _{ij})^{2}/(4a^2)$. 
Then, the second-order action for tensor perturbations 
in GLPV theory reads \cite{Piazza,Tsuji14}
\begin{eqnarray}
S_2^{(h)} &=&
\int d^4 x\,a^{3}\left[ L_{,\s} \left( \delta K_{\mu
}^{\nu}\delta K_{\nu }^{\mu}-\delta K^{2}\right) +
\mathcal{E}\delta _{2} 
\mathcal{R}\right] \nonumber \\
&=&\int d^4 x\,\frac{a^{3}}{4} \left[
L_{,\s} \dot{\gamma}_{ij}^2-\frac{{\cal E}}{a^2}
(\partial_k \gamma_{ij})^2 \right]\,.
\label{tenlag}
\end{eqnarray}

Then, the conditions for the absence of 
tensor ghosts and Laplacian instabilities 
in GLPV theory read
\begin{eqnarray}
L_{,\s} &=& -A_4-3HA_5>0\,,\label{tenno}\\
{\cal E} &=& B_4+\frac12 \dot{B}_5>0\,.
\label{tenpro}
\end{eqnarray}
Recall that in Horndeski theory the explicit forms of 
$L_{,\s}$ and ${\cal E}$ are given by 
Eqs.~(\ref{Ls2}) and (\ref{Eho}), respectively. 
The no-ghost condition (\ref{tenno}) does not 
involve the dependence of $B_4$ and $B_5$, 
but the condition (\ref{tenpro}), which is related to 
the tensor propagation speed square 
$c_t^2={\cal E}/L_{,\s}$, depends on $B_4$ and $B_5$.
Hence the scalar and tensor propagation speed squares 
are important quantities to distinguish between the theories 
with same values of $A_2,A_3,A_4,A_5$ but with different 
values of $B_4,B_5$.

Under the condition (\ref{tenno}) the no-ghost condition 
$Q_s>0$ of scalar perturbations translates to 
\be
9{\cal W}^2+8L_{,\s} w>0\,.
\ee
We also note that, under the condition (\ref{tenpro}), the quantity 
$(a{\cal M})^{\cdot}$ must be positive to realize $c_s^2>0$ 
in Eq.~(\ref{cs}). 

The above results are consistent with those derived by 
direct variation of the original Horndeski Lagrangian (\ref{Lagsum}) 
with (\ref{eachlag2})-(\ref{L5lag}).
The quantities $w_1,w_2,w_3,w_4$ introduced in Ref.~\cite{infnon}
have the correspondence 
$w_1 \to 2L_{,\s}^{{\rm (H)}}$, 
$w_2 \to {\cal W}^{{\rm (H)}}$, 
$w_3 \to w^{{\rm (H)}}$, and 
$w_4 \to 2{\cal E}^{{\rm (H)}}$ 
(with the replacement $X \to -2X$, $G_2 \to P$, $G_3 \to -G_3$, and 
$2G_4 \to M_{\rm pl}^2F$ to recover Eqs.~(18)-(21) of Ref.~\cite{infnon}).

\subsection{The inflationary power spectra of curvature 
and tensor perturbations}

The scalar degree of freedom appearing in Horndeski and 
GLPV theories can be responsible for inflation in the early Universe. 
This is possible if the field $\phi$ evolves slowly along a nearly 
flat potential $V(\phi)$ (slow-roll inflation \cite{newinf,chaotic}) 
or if the presence of higher-order field kinetic terms gives rise to a fixed point 
characterized by a nearly constant kinetic energy (k-inflation \cite{kinf}).
In both cases the Hubble parameter $H$ is nearly constant 
during the inflationary period, so the slow-roll parameter 
defined by 
\be
\epsilon \equiv -\frac{\dot{H}}{H^2}
\ee
is much smaller than 1. 
We assume that the terms without containing the scale factor $a$ 
evolve slowly during inflation, so that the quantities
\begin{equation}
\delta_{Q_s} \equiv \frac{\dot{Q}_s}{HQ_s}\,,\qquad
\delta_{c_s}  \equiv \frac{\dot{c}_s}{Hc_s}
\label{Qcdef}
\end{equation}
are much smaller than unity.

The curvature perturbation $\zeta$ generated from quantum 
fluctuations in the early Universe can be responsible for the origin of 
observed CMB temperature anisotropies \cite{oldper}.
Let us derive the primordial power spectrum of $\zeta$ 
generated during inflation.
For this purpose we express $\zeta$ in Fourier space, as 
\begin{equation}
\zeta (\tau,{\bm{x}})
= \frac{1}{(2\pi)^{3}}\int d^{3}{\bm{k}}\, 
\hat{\zeta} (\tau,{\bm{k}})e^{i{\bm{k}}\cdot{\bm{x}}}\,,
\label{RFourier}
\end{equation}
where $\tau \equiv \int a^{-1}\,dt$ is the conformal time,
${\bm k}$ is the comoving wavenumber, and
\begin{equation}
\hat{\zeta}(\tau,{\bm{k}})=u(\tau,{\bm{k}})a({\bm{k}})
+u^{*}(\tau,{-\bm{k}})a^{\dagger}(-{\bm{k}})\,.
\end{equation}
The annihilation operator $a({\bm{k}})$ and the creation 
operator $a^{\dagger}({\bm{k}})$ obey the commutation relations
\begin{eqnarray}
& &
\left[a({\bm{k}}_{1}),a^{\dagger}({\bm{k}}_{2})\right]
=(2\pi)^{3}\delta^{(3)}({\bm{k}}_{1}-{\bm{k}}_{2})\,,\nonumber \\
& &
\left[a({\bm{k}}_{1}),a({\bm{k}}_{2})\right]
=\left[a^{\dagger}({\bm{k}}_{1}),a^{\dagger}({\bm{k}}_{2})\right]=0\,.
\end{eqnarray}

In Horndeski and GLPV theories the second-order Lagrangian density 
for $\zeta$ is given by Eq.~(\ref{perlag}).
Defining a rescaled field $v=zu$ with $z=a\sqrt{2Q_s}$,
the kinetic term in the second-order action 
$S_2=\int d^4x\,{\cal L}_2$
can be expressed as $\int d\tau d^3 x\,v'^2/2$, 
where a  prime represents a derivative with respect to $\tau$. 
Hence $v$ corresponds to a canonical field associated with 
the quantization procedure. 
{}From Eq.~(\ref{pereq}) the field $v(\tau,{\bm{k}})$ 
obeys the equation of motion 
\begin{equation}
v''+\left(c_{s}^{2}k^{2}-\frac{z''}{z}\right)v=0\,.
\label{veq}
\end{equation}

Since $H$ is nearly constant during inflation, it follows that 
$\tau \simeq -1/(aH)$ (where the integration constant is set to 0). 
As long as $Q_s$ is nearly constant (i.e., $|\delta_{Q_s}| \ll 1$), 
the quantity $z''/z$ in Eq.~(\ref{veq}) is approximately 
given by $z''/z \simeq 2/\tau^{2}$.
If we go back to the asymptotic past ($\tau \to -\infty$), 
Eq.~(\ref{veq}) reduces to $v''+c_{s}^{2}k^{2}v \simeq 0$.
Choosing the Bunch-Davies vacuum in this limit,
the solution to this equation is given by 
$v=e^{-ic_{s}k\tau}/\sqrt{2c_{s}k}$ for $\tau \to -\infty$.
The term $z''/z$ characterizes the effect of gravity, which 
becomes comparable to $c_s^2k^2$ for $c_s k \approx aH$. 
Since the gravitational term dominates over $c_s^2k^2$
in the regime $c_sk \ll aH$, the solution to Eq.~(\ref{veq}) 
is given by $v \propto z$, i.e., $u=$\,constant.
In other words, the Fourier components of $\zeta$ are 
``frozen'' for $c_sk<aH$. 

More precisely, the solution to Eq.~(\ref{veq}) on the 
de Sitter background recovering 
the Bunch-Davies vacuum in the asymptotic past is given by
\begin{equation}
u (\tau, k)=\frac{i\,H\, e^{-ic_{s}k\tau}}
{2(c_{s}k)^{3/2}\sqrt{Q_s}}\,(1+ic_{s}k\tau)\,.
\label{usol}
\end{equation}
Strictly speaking, the Hubble parameter varies during inflation, but 
its effect appears only as a next-order 
slow-roll correction to the power spectrum \cite{Chen}.
We are interested in the two-point correlation function 
of $\zeta$ in the regime $c_sk \ll aH$, i.e., 
the vacuum expectation value
$\langle 0| \hat{\zeta} (\tau, {\bm k}_1) \hat{\zeta} 
(\tau,{\bm k}_2) | 0 \rangle$ at $\tau \approx 0$. 
We define the scalar power spectrum 
${\cal P}_{\zeta} (k_1)$, as 
\begin{equation}
\langle 0| \hat{\zeta} (0,{\bm k}_1) \hat{\zeta} (0,{\bm k}_2) | 0 \rangle
=\frac{2\pi^2}{k_1^3} {\cal P}_{\zeta} (k_1)\,
(2\pi)^3 \delta^{(3)} ({\bm k}_1+{\bm k}_2)\,.
\label{powerdef}
\end{equation}

On using the solution (\ref{usol}), we obtain
\begin{equation}
{\cal P}_{\zeta}=\frac{H^2}{8\pi^2 Q_s c_s^3}\,.
\label{scalarpower}
\end{equation}
Since the curvature perturbation is frozen in the regime 
$c_s k<aH$, we can compute
the power spectrum (\ref{scalarpower}) at
$c_s k=aH$ during inflation. 
The result (\ref{scalarpower}) matches with that derived in 
Horndeski theory \cite{KYY,Steer,infnon}.
{}From the Planck data the scalar amplitude is 
constrained to be ${\cal P}_{\zeta} \simeq  2.2 \times 10^{-9}$ 
for the wavenumber $k_0=0.002$ Mpc$^{-1}$ \cite{Planck}.
We define the spectral index of ${\cal P}_\zeta$, as
\begin{eqnarray}
n_{s}-1 \equiv \frac{d \ln {\cal P}_{\zeta}}
{d \ln k}\bigg|_{c_sk=aH}
\simeq -2\epsilon-\delta_{Q_s}-3\delta_{c_s}\,.
\label{ns}
\end{eqnarray}
Since the slow-roll parameters $\epsilon, \delta_{Q_s}, \delta_{c_s}$
are much smaller than 1, the power spectrum ${\cal P}_{\zeta}$ 
is close to scale-invariant ($n_s \simeq 1$).
Since the deviation of $n_s$ from 1 is different depending on 
the models of inflation, we can distinguish them from precise 
measurements of the CMB temperature anisotropies \cite{Planckinf,Kuro}. 
Assuming that the running spectral index 
$\alpha_s=dn_s/\ln k|_{c_sk=aH}$ is negligible, the Planck 
data put the constraint $n_s=0.9603 \pm 0.0073$ at 
68 \% confidence level \cite{Planck}.

For gravitational waves we need to express the tensor perturbation 
$\gamma_{ij}$ in terms of the two polarization modes 
$e_{ij}^{+}$ and $e_{ij}^{\times}$, as 
$\gamma_{ij}=h_+e_{ij}^{+}+h_{\times}e_{ij}^{\times}$.
In Fourier space, the polarization tensors 
satisfy the normalization condition
$e_{ij} (\bm{k})\,e_{ij} (-\bm{k})^{*}=2$ for each 
mode and the orthogonality condition 
$e^{+}_{ij} (\bm{k})\,e^{\times}_{ij}(-\bm{k})^{*}=0$. 
Then, the second-order action (\ref{tenlag}) reads 
\be
S_2^{(h)} =
\sum_{\lambda=+,\times}
\int d^4 x~a^{3} Q_t \left[ \dot{h}_{\lambda}^2
-\frac{c_t^2}{a^2} (\partial h_{\lambda})^2 \right]\,,
\label{tenlag2}
\ee
where
\be
Q_t \equiv \frac{L_{,\s}}{2}\,,\qquad 
c_t^2 \equiv \frac{\mathcal{E}}{L_{,\s}}\,.
\ee
Following the similar procedure to that for scalar perturbations, 
it is straightforward to derive the power spectrum 
${\cal P}_h$ of gravitational waves \cite{Tsuji14}.
In the regime $c_tk<aH$ the tensor perturbation is frozen, so the 
resulting power spectrum is given by 
\be
{\cal P}_h=\frac{H^2}{2\pi^2 Q_tc_t^3}\,,
\label{tensorpower}
\ee
which should be evaluated at $c_tk=aH$.

{}From Eqs.~(\ref{scalarpower}) and (\ref{tensorpower}) the tensor-to-scalar 
ratio reads
\be
r \equiv \frac{{\cal P}_h}{{\cal P}_{\zeta}}
=4\frac{Q_s c_s^3}{Q_t c_t^3}\,,
\label{rfi}
\ee
where we have neglected the difference for the moments at which 
scalar and tensor perturbations are frozen (which appears as the next-order 
slow-roll correction).
The combined analysis of the Planck data with the WMAP large-angle polarization 
measurement and ACT/SPT temperature data put the constraint 
$r<0.11$ at 95 \% confidence level \cite{Planck}.
The scalar spectral index $n_s$ and the tensor-to-scalar ratio $r$ 
are two key quantities to distinguish between many inflationary models.
See Refs.~\cite{Kuro} for detailed constraints on the inflationary models 
in the framework of Horndeski theory.

\subsection{Dark energy in the presence of matter}
\label{DEsec}

Let us consider the application of Horndeski and GLPV theories 
to dark energy. We assume that the scalar degree of freedom $\phi$ 
is responsible for the late-time cosmic acceleration.
In order to discuss the cosmological dynamics associated with dark energy, 
we need to incorporate other sources of matter such as dark matter, 
baryons, and radiation. For this purpose, we take into account 
k-essence type matter described by the Lagrangian 
$P(\varphi,Y)$ \cite{Scherrer,Daniel,Laszlo,Gleyzes}, 
where $P$ is an arbitrary function of another scalar field $\varphi$ 
and its kinetic energy $Y=g^{\mu \nu} \partial_{\mu} \varphi \partial_{\nu} 
\varphi$.
A more general system with multiple scalars 
(to accommodate both non-relativistic matter and radiation) 
has been studied in detail in Ref.~\cite{Kase}.

The action of Horndeski and GLPV theories with k-essence matter 
is described by 
\be
S=\int d^4 x \sqrt{-g} \left[ L(N,K,{\cal S},{\cal R},{\cal U};t)
+P(\varphi,Y) \right]\,.
\label{actionma}
\ee
The equations of motion for the background and linear perturbations 
can be derived by expanding the action (\ref{actionma}) up to second order 
in perturbations.
Note that the unitary gauge $\delta \phi=0$ is chosen 
for the dark energy field $\phi$. 

Using the fact that the first-order perturbation of $Y$ is given by 
\be
\delta_1 Y=2\dot{\varphi}^2 \delta N-2\dot{\varphi} \dot{\delta \varphi}\,,
\label{del1Y}
\ee
the first-order Lagrangian density (\ref{L1}) is modified to
\be
\mathcal{L}_{1}=
a^{3} ( \bar{L}+L_{,N}-3H\mathcal{F}
+P+2P_{,Y} \dot{\varphi}^2 ) \delta N
+( \bar{L}-\dot{\mathcal{F}}-3H\mathcal{F}+P) \delta \sqrt{h}
+a^3 P_{,\varphi} \delta \varphi-2a^3P_{,Y} 
\dot{\varphi} \dot{\delta \varphi}
+a^3\mathcal{E} \delta_1 {\cal R}\,.
\label{L1m}
\ee
Varying the Lagrangian density (\ref{L1m}) with respect to 
$\delta N$, $\delta \sqrt{h}$, $\delta \varphi$, 
it follows that 
\ba
& &
\bar{L}+L_{,N}-3H\mathcal{F}=\rho_m\,,
\label{backma1} \\
& & 
\bar{L}-\dot{\mathcal{F}}-3H\mathcal{F}=-P_m\,,
\label{backma2} \\
& &
\frac{d}{dt} \left( a^3 P_{,Y} \dot{\varphi} \right)
+\frac12a^3 P_{,\varphi}=0\,,
\label{scalareq}
\ea
where $\rho_m$ and $P_m$ are energy density 
and pressure of the scalar field $\varphi$, 
defined, respectively, by 
\be
\rho_m \equiv -P-2P_{,Y} \dot{\varphi}^2\,,\qquad
P_m \equiv P\,.
\label{rhom}
\ee
In terms of $\rho_m$ and $P_m$, the scalar field 
equation of motion (\ref{scalareq}) can be written 
in the form
\be
\dot{\rho}_m+3H(\rho_m+P_m)=0\,,
\label{continuity}
\ee
which corresponds to the standard continuity equation 
of matter.

In what follows we assume that the field Lagrangian 
depends on $Y$ alone, i.e., 
\be
P=P(Y)\,.
\ee
In this case the scalar field $\varphi$ behaves as a 
barotropic perfect fluid \cite{Hu}. 
Expanding the action (\ref{actionma}) up to second order in perturbations, 
we find that the following term is added to 
the Lagrangian density (\ref{L2den}) : 
\be
{\cal L}_2^{M} \equiv 
P_{,Y} \delta \sqrt{h}\,\delta_1 Y+a^3 
\left( P_{,Y} \delta_2 Y+P_{,YY} \delta_1 Y^2/2
+P_{,Y}\delta N \delta_1 Y \right)\,,
\ee
where the second-order contribution to $Y$ 
is given by 
\be
\delta_2 Y=-\dot{\delta \varphi}^2-3\dot{\varphi}^2 \delta N^2
+4\dot{\varphi} \dot{\delta \varphi} \delta N
+2\dot{\varphi} \partial_j \psi \partial^j \delta \varphi
+\frac{1}{a^2}(\partial \delta \varphi)^2 \,.
\ee

Using the fact that both Horndeski and GLPV theories satisfy 
the conditions (\ref{nospa}), the second-order Lagrangian 
density reads \cite{Kase}
\ba
\hspace{-0.5cm}
{\cal L}_2 &=& a^3 \biggl\{ \frac12 (2L_{,N}+L_{,NN}-6H{\cal W}
+12H^2L_{,\cal S}) \delta N^2+
\left[ {\cal W} ( 3\dot{\zeta}-\Delta \psi)
-4 ({\cal D}+{\cal E}) \Delta \zeta
\right] \delta N +4L_{,\cal S} \dot{\zeta} \Delta \psi
-6L_{,\cal S} \dot{\zeta}^2
\nonumber \\
\hspace{-0.5cm}
& &
+2{\cal E} \frac{(\partial \zeta)^2}{a^2} 
+(2\dot{\varphi}^2 P_{,YY}
-P_{,Y}) (\dot{\varphi}^2 \delta N^2-2\dot{\varphi}
\dot{\delta \varphi} \delta N+\dot{\delta \varphi}^2)
-6\dot{\varphi} P_{,Y} \zeta \dot{\delta \varphi} 
-2\dot{\varphi} P_{,Y} \delta \varphi
\Delta \psi
+P_{,Y} \frac{(\partial \delta \varphi)^2}{a^2}
\biggr\}\,.
\label{L2f}
\ea
Variations of this Lagrangian density with respect to 
$\delta N$ and $\psi$ lead to the following Hamiltonian 
and momentum constraints respectively:
\ba
\hspace{-1cm}
& & 
(2L_{,N}+L_{,NN}-6H{\cal W}+12H^2L_{,\cal S}) \delta N
+ {\cal W} ( 3\dot{\zeta}-\Delta \psi)
-4({\cal D}+{\cal E}) \Delta \zeta
+2\dot{\varphi} 
(P_{,Y}-2\dot{\varphi}^2 P_{,YY}) 
(\dot{\delta \varphi}-\dot{\varphi} \delta N)=0\,,\label{Hamima}\\
\hspace{-1cm}
& & {\cal W} \delta N-4L_{,\cal S} \dot{\zeta}
+2\dot{\varphi} P_{,Y} \delta \varphi=0\,.
\label{momenma}
\ea
We solve Eqs.~(\ref{Hamima}) and (\ref{momenma}) for $\delta N$, 
$\Delta \psi$ and then substitute them into Eq.~(\ref{L2f}).
The resulting Lagrangian density is expressed in 
the following form
\be
\mathcal{L}_{2}=a^{3}\left( \dot{\vec{\mathcal{X}}}^{t}{\bm K} 
\dot{\vec{\mathcal{X}}}-\partial _{j}\vec{\mathcal{X}}^{t}{\bm G}
\partial^{j}{\vec{\mathcal{X}}}-\vec{\mathcal{X}}^{t}{\bm B} 
\dot{\vec{\mathcal{X}}}-\vec{\mathcal{X}}^{t}{\bm M} \vec{\mathcal{X}}\right) \,,
\label{L2mat}
\ee
where ${\bm K}$, ${\bm G}$, ${\bm B}$, ${\bm M}$ 
are $2 \times 2$ matrices, and the dimensionless vector 
$\vec{\mathcal{X}}$ is defined by
\be
\vec{\mathcal{X}}^{t}=\left( \zeta, \delta \varphi /M_{\mathrm{pl}} 
\right) \,.
\ee
The no-ghost conditions and the scalar propagation speeds are 
determined by the two matrices ${\bm K}$ and ${\bm G}$, 
whose components are 
\ba
& &
K_{11}=Q_s+\frac{16L_{,\s}^2}{M_{\rm pl}^2{\cal W}^2} 
\dot{\varphi}^2 K_{22}\,,\qquad
K_{22}=(2\dot{\varphi}^2P_{,YY}-P_{,Y})M_{\rm pl}^2\,,
\qquad
K_{12}=K_{21}=-\frac{4L_{,\s}\dot{\varphi}}{M_{\rm pl}{\cal W}}
K_{22}\,,\\
& & 
G_{11}=2(\dot{\cal M}+H{\cal M}-{\cal E})\,,\qquad
G_{22}=-P_{,Y}M_{\rm pl}^2\,,\qquad
G_{12}=G_{21}=-\frac{{\cal M}\dot{\varphi}}{L_{,\s}M_{\rm pl}}G_{22}\,,
\ea
where $Q_s$ and ${\cal M}$ are defined by Eqs.~(\ref{Qsdef}) and 
(\ref{Mdef}), respectively.

Provided that the symmetric matrix ${\bm K}$ is positive definite, 
the scalar ghosts are absent.
The positivity of ${\bm K}$ translates to the conditions 
that the determinants of principal sub-matrices 
of ${\bm K}$ are positive \cite{Kase}, i.e., 
\ba
& &
Q_s+\frac{16L_{,\s}^2}{M_{\rm pl}^2{\cal W}^2} 
\dot{\varphi}^2 K_{22}>0\,,\\
& &
Q_sK_{22}>0\,.
\ea
These conditions are satisfied for $Q_s>0$ and $K_{22}>0$.

In the limit of large wavenumber $k$, the Lagrangian density (\ref{L2mat}) 
gives rise to the dispersion relation 
\be
{\rm det} \left( \omega^2 {\bm K}-k^2{\bm G}/a^2 
\right)=0\,,
\label{cseq}
\ee
where $\omega$ is a frequency. 
Defining the scalar propagation speed $c_s$ as 
$\omega^2=c_s^2\,k^2/a^2$, Eq.~(\ref{cseq}) translates to
\be
\left(c_s^2 K_{11}-G_{11} \right)
\left(c_s^2 K_{22}-G_{22} \right)
-\left(c_s^2 K_{12}-G_{12} \right)^2=0\,.
\label{cseq2}
\ee
In Horndeski theory we have the particular relation 
${\cal M}=4L_{,\s}^2/{\cal W}$, see Eq.~(\ref{Mre}). 
Then, the propagation speed $c_{s\rm H}$ satisfies
\be
c_{s\rm H}^2 K_{12}-G_{12}=
-\frac{4L_{,\cal S} \dot{\varphi}}{M_{\rm pl}{\cal W}}
\left( c_{s\rm H}^2 K_{22} -G_{22} \right)\,,
\label{cHre}
\ee
where the lower index ``H'' represents the values in 
Horndeski theory.
Substituting Eq.~(\ref{cHre}) into Eq.~(\ref{cseq2}), we obtain 
the following two solutions 
\ba
& &
c_{s{\rm H}1}^2=\frac{G_{11} 
- [4L_{,\cal S}\dot{\varphi}/(M_{\rm pl}{\cal W})]^2 
G_{22}}{K_{11} 
-[4L_{,\cal S}\dot{\varphi}/(M_{\rm pl}{\cal W})]^2 
K_{22}}=
\frac{1}{Q_s} \left[ 
2(\dot{\cal M}+H{\cal M}-{\cal E})
+\left( \frac{4L_{,\s}\dot{\varphi}}{{\cal W}} 
\right)^2 P_{,Y} \right]\,,\label{cH1}\\
& & 
c_{s{\rm H}2}^2=\frac{G_{22}}{K_{22}}
=\frac{P_{,Y}}
{P_{,Y}-2\dot{\varphi}^2
P_{,YY}}\,.
\label{cH2}
\ea
Compared to the result (\ref{cs}), the sound speed square
$c_{s{\rm H}1}^2$ of the dark energy field $\phi$ 
is modified by the presence of an additional scalar field $\varphi$. 
Meanwhile, the sound speed square $c_{s{\rm H}2}^2$ 
of k-essence matter (originally derived in Ref.~\cite{Garriga}) 
is not affected by the dark energy field
in Horndeski theory.

In GLPV theory the relation (\ref{cHre}) no longer holds, so 
the solutions to Eq.~(\ref{cseq2}) are not given by 
Eqs.~(\ref{cH1}) and (\ref{cH2}).
In this case, we can express Eq.~(\ref{cseq2}) in the 
following form
\be
\left( c_s^2-c_{s{\rm H}1}^2 \right) \left( c_s^2-c_{s{\rm H}2}^2 \right)
=\frac{16L_{,\s}^2}{Q_s{\cal W}^2} \left( \frac{{\cal M}{\cal W}}{4L_{,\s}^2} 
-1 \right) \dot{\varphi}^2 P_{,Y} 
\left[ 2c_s^2-c_{s{\rm H}2}^2 \left( \frac{{\cal{M}}{\cal W}}{4L_{,\s}^2}+1
\right) \right]\,.
\label{ceqN=2}
\ee
In Horndeski theory the r.h.s. of Eq.~(\ref{ceqN=2}) vanishes 
due to the second relation of Eq.~(\ref{Mre}), 
so we reproduce the solutions (\ref{cH1}) and (\ref{cH2}).
In GLPV theory both $c_{s{\rm H}1}^2$ and $c_{s{\rm H}2}^2$ 
are modified by the presence of the dark energy field.
This is the important difference to distinguish between 
the two theories.

For example, let us consider the difference between covariant 
and covariantized Galileons discussed in Sec.\,\ref{basec}. 
In both theories the functions $A_2, A_3, A_4, A_5$ appearing 
in the Lagrangian (\ref{LH2}) are given by (\ref{cova1}), so 
Eqs.~(\ref{backma1}) and (\ref{backma2}) read
\ba
\hspace{-1.2cm}
& &
3M_{\rm pl}^2 H^2=
-\frac12 c_2 \dot{\phi}^2+\frac{3c_3H \dot{\phi}^3}{M^3}
-\frac{45c_4H^2 \dot{\phi}^4}{2M^6}
+\frac{21c_5H^3\dot{\phi}^5}{M^9}+\rho_m\,,
\label{backeq1} \\
\hspace{-1.2cm}
& &
3M_{\rm pl}^2 H^2+2M_{\rm pl}^2 \dot{H}
=\frac12 c_2 \dot{\phi}^2+\frac{c_3\dot{\phi}^2 \ddot{\phi}}
{M^3}-\frac{3c_4 \dot{\phi}^3}{2M^6}
\left[ 8H \ddot{\phi}+(3H^2+2\dot{H})\dot{\phi} \right]
+\frac{3c_5H \dot{\phi}^4}{M^9}
\left[ 5H \ddot{\phi} +2(H^2+\dot{H}) \dot{\phi} \right]
-P_m\,.
\label{backeq2}
\ea
These equations show that there exists a de Sitter solution 
characterized by $H=$\,constant and $\dot{\phi}=$\,constant 
with $\rho_m=P_m=0$.
In fact, this de Sitter solution can be used for realizing 
the late-time cosmic acceleration \cite{GS,Gacosmo}.

The perfect fluids of radiation and non-relativistic matter 
can be accommodated by considering the Lagrangians 
$P(Y)=b_1Y^2$ and $P(Y)=b_2(Y-Y_0)^2$ with 
$|Y-Y_0| \ll Y_0$, respectively 
($b_1,b_2,Y_0$ are constants) \cite{Scherrer,Kase}.
During the radiation and matter eras
the background equations of motion 
allow the existence of tracker 
solutions characterized by $H\dot{\phi}={\rm constant}$ \cite{Gacosmo}. 
Along the tracker, each derivative term on the r.h.s. 
of Eqs.~(\ref{backeq1}) and (\ref{backeq2}) is proportional 
to $H^{-1}$. During the cosmological sequence of radiation, 
matter, and de Sitter epochs, the dark energy equation of 
state evolves as $w_{\rm DE}=-7/3 \to -2 \to -1$.
However, this evolution is in tension with the 
joint data analysis of SN Ia, CMB, 
and baryon acoustic oscillations \cite{Savvas}.

The solutions approaching the tracker at late times can be 
compatible with the observational data. 
For the late-time tracker, the quantity
$r_1 \equiv H_{\rm dS} \dot{\phi}_{\rm dS}/(H \dot{\phi})$
(where ``dS'' represents the values at the late-time 
de Sitter solution)
is initially much smaller than 1 and then $r_1$ approaches 
the order of unity only recently.

In the covariant Galileon, which belongs to a class of Horndeski theory, 
the scalar propagation speed square (\ref{cH1}) is given by 
$c_{s{\rm H}1}^2=(\Omega_r+1)/40$ in the regime $r_1 \ll 1$ \cite{Gacosmo}, 
where $\Omega_r$ is the density parameter of radiation. 
Since $c_{s{\rm H}1}^2=1/20$ and $1/40$ during the 
radiation and matter eras respectively, the Laplacian 
instabilities of the dark energy perturbation are absent 
in these epochs. Note that the matter sound speed 
(\ref{cH2}) is given by $c_{s{\rm H}2}^2=1/3$ for 
radiation and $c_{s{\rm H}2}^2 \simeq +0$ for 
non-relativistic matter. 
The evolution of matter density perturbations and
observational tests for the covariant Galileon with 
large-scale structure data have been studied 
in Refs.~\cite{Galima}.

In the covariantized Galileon, which goes beyond the realm of Horndeski theory,
the two propagation speeds $c_s$ are known by solving Eq.~(\ref{ceqN=2}). 
The detailed analysis in Ref.~\cite{Kase} showed that the matter sound 
speed square $c_{s2}^2$ is close to the value $c_{s{\rm H}2}^2$, 
but the dark energy sound speed square $c_{s1}^2$ 
of the late-time tracking solution is modified as $c_{s1}^2=(3\Omega_r-1)/40$ 
in the regime $r_1 \ll 1$.
Since $c_{s1}^2=-1/40$ during the matter era, the dark energy model 
based on the covariantized Galileon is plagued by small-scale 
Laplacian instabilities. 
In spite of the fact that the two Galileon theories give rise to 
the same background equations of motion on the flat FLRW background, 
they are clearly distinguished by the scalar propagation speeds.

\section{Application to Ho\v{r}ava-Lifshitz gravity}
\label{apsec2} 

We apply the results in Sec.~\ref{persec} to Ho\v{r}ava-Lifshitz gravity.
In what follows we shall consider the projectable and non-projectable 
versions of the theory separately.

\subsection{Projectable Ho\v{r}ava-Lifshitz gravity}

The projectable version \cite{Horava} corresponds to the case where 
the lapse $N$ is a function of $t$ alone.
In this case all the terms $\alpha_i$ ($i=1,2,\cdots,5$) defined in 
Eq.~(\ref{aldef}), which come from the acceleration 
$a_i=\nabla_i \ln N$, vanish. 
Then, the theory is described by the action
\be
S=\int d^4 x \sqrt{-g} \left[ L(K,{\cal S},{\cal R},{\cal Z},
{\cal Z}_1,{\cal Z}_2)+L_m \right]\,,
\label{Horava1}
\ee
where 
\be
L =
\frac{M_{\rm pl}^2}{2} \left[ {\cal S}-\lambda K^2
+{\cal R}
-M_{\rm pl}^{-2} \left( g_2 {\cal R}^2+g_3 {\cal Z} \right)
-M_{\rm pl}^{-4}\left( g_4 {\cal Z}_1+g_5 {\cal Z}_2
\right) \right]\,,
\ee
and $L_m$ is a matter Lagrangian.

To derive the background equations of motion, 
we consider the k-essence matter Lagrangian 
$L_m=P(\varphi,Y)$ introduced in Sec.\,\ref{DEsec}. 
Note that the matter sector can potentially couple to the gravitational sector 
at high energy, but we simply assume the absence of such couplings 
in the following discussion.

Since $N=N(t)$ in the present case, the Lagrangian 
(\ref{Horava1}) cannot be varied with respect to $\delta N$ 
to derive one of the background equations of motion. 
Varying the action (\ref{Horava1}) with respect to $\delta \sqrt{h}$, 
we obtain the same form of equation as (\ref{backma2}) with 
$\bar{L}=3M_{\rm pl}^2(1-3\lambda)H^2/2$, 
${\cal F}=M_{\rm pl}^2 (1-3\lambda)H$, and 
the pressure $P_m=P$, i.e., 
\be
\frac{3\lambda-1}{2}M_{\rm pl}^2 (2\dot{H}+3H^2)=-P_m\,.
\label{Hoback2}
\ee
Variation of the action with respect to $\varphi$ leads to 
the same equation as (\ref{scalareq}):
\be
\frac{d}{dt} \left( a^3 P_{,Y} \dot{\varphi} \right)
+\frac12a^3 P_{,\varphi}=0\,.
\label{varphi}
\ee
Defining the field energy density $\rho_m=-P-2P_{,Y} \dot{\varphi}^2$ 
as Eq.~(\ref{rhom}), Eq.~(\ref{varphi}) translates to 
the continuity equation (\ref{continuity}), i.e.,
\be
P_m=-\frac{1}{3a^3H} \frac{d}{dt} (a^3 \rho_m)\,.
\label{Hoback3}
\ee
Integrating Eq.~(\ref{Hoback2}) with respect to $t$ 
after substitution of Eq.~(\ref{Hoback3}), we obtain 
\be
\frac32 (3\lambda-1)M_{\rm pl}^2H^2
=\rho_m +\frac{C}{a^3}\,,
\label{Hoback1}
\ee
where $C$ is an integration constant. 
The extra term $C/a^3$ behaves as non-relativistic dark matter \cite{Mukoh}. 
In the projectable version of Ho\v{r}ava-Lifshitz gravity, such apparent 
matter arises due to the absence of the Hamiltonian constraint.

Let us proceed to the discussion of cosmological perturbations 
in the absence of matter ($L_m=0$).
The projectable version of Ho\v{r}ava-Lifshitz gravity 
is characterized by the gauge choice
\be
\delta N=0\,,
\label{delNga}
\ee
which is consistent with the foliation-preserving transformation
$t \to t+f(t)$. We cannot employ the perturbation 
equation of motion (\ref{Hami}) that corresponds 
to the Hamiltonian constraint.
Since ${\cal A}=-\lambda M_{\rm pl}^2$ and 
$L_{,\s}=M_{\rm pl}^2/2$, ${\cal W}=(3\lambda-1)M_{\rm pl}^2 H$, 
and ${\cal C}=0$, Eq~(\ref{momen}) gives the relation 
\be
\Delta \psi=\frac{3\lambda-1}{\lambda-1} \dot{\zeta}\,,
\label{delpsiga}
\ee
for $\lambda \neq 1$. 
Then, the quantity ${\cal Y}$ defined in Eq.~(\ref{Y2}) is given by 
${\cal Y}=2M_{\rm pl}^2 (3\lambda-1)\dot{\zeta}/(\lambda-1)$.
On using the relations ${\cal D}=0$, ${\cal E}=M_{\rm pl}^2/2$, 
$4{\cal G}+3L_{,{\cal Z}}=-(8g_2+3g_3)/2$, and 
$8L_{,{\cal Z}_1}+3L_{,{\cal Z}_2}=-(8g_4+3g_5)/(2M_{\rm pl}^2)$, 
Eq.~(\ref{zeta}) reads
\be
\frac{d}{dt} \left( \frac{3\lambda-1}{\lambda-1} a^3 \dot{\zeta} 
\right)+a^3{\cal O} \zeta=0\,,
\label{zetaho}
\ee
where 
\be
{\cal O} \equiv \Delta+\frac{\Delta^2}{M_2^2}-\frac{\Delta^3}{M_3^4}\,,
\label{ope}
\ee
and 
\be
M_2^2 \equiv M_{\rm pl}^2 (8g_2+3g_3)^{-1}\,,\qquad
M_3^4 \equiv M_{\rm pl}^4 (8g_4+3g_5)^{-1}\,.
\label{M23}
\ee
Substituting Eqs.~(\ref{delNga}) and (\ref{delpsiga}) into 
the second-order Lagrangian density (\ref{L2exp}), 
we also obtain
\be
{\cal L}_2=M_{\rm pl}^2 a^3 \left( \frac{3\lambda-1}{\lambda-1} 
\dot{\zeta}^2-\zeta {\cal O} \zeta \right)\,,
\label{lagho}
\ee
which matches with the results in 
Refs.~\cite{strong2,Mukoreview} 
(see also Refs.~\cite{Hocosmo,Wang}).
In fact, variation of ${\cal L}_2$ with respect to $\zeta$ 
leads to the equation of motion (\ref{zetaho}).
 
The breaking of gauge symmetry in GR ($\lambda \neq 1$) 
gives rise to the propagation of the scalar degree of freedom 
$\zeta$. In order to evade the appearance of scalar 
ghosts, we require that $(3\lambda-1)/(\lambda-1)>0$, i.e., 
$\lambda>1$ or $\lambda<1/3$. 
At low energy we should recover the behavior similar to GR, 
so it is natural to focus on the regime $\lambda>1$.

In Minkowski space-time ($a=1$),
we obtain the following dispersion relation from 
the Lagrangian density (\ref{lagho}):
\be
\omega^2=\frac{\lambda-1}{3\lambda-1} 
\left( \frac{k^6}{M_3^4}+\frac{k^4}{M_2^2}
-k^2 \right)\,.
\label{dispersion}
\ee
For the wavenumber $k$ much smaller than 
$M_3$ and $M_2$ this relation reduces to 
$\omega^2 \simeq -(\lambda-1)k^2/(3\lambda-1)$, 
so the scalar propagation speed square 
$c_s^2=\omega^2/k^2=-(\lambda-1)/(3\lambda-1)$ 
is negative under the no-ghost condition 
$(3\lambda-1)/(\lambda-1)>0$.
The time scale associated with this Laplacian 
instability can be estimated as
\be
t_L \approx \frac{1}{k} \sqrt{\frac{3\lambda-1}{\lambda-1}}\,.
\ee

On the cosmological background there is a Hubble friction 
term $3H\dot{\zeta} (3\lambda-1)/(\lambda-1)$ appearing 
in Eq.~(\ref{zetaho}). 
Provided that this term dominates over the Laplacian term 
$-k^2 \zeta$, it is possible to avoid the instability of 
scalar perturbations. This translates to the condition 
$t_L \gg H^{-1}$. 
We also note that the time scale associated the growth 
of large-scale structures in the Universe is given by 
$t_J \approx M_{\rm pl}/\sqrt{\rho}$, where $\rho$ is 
the energy density of non-relativistic matter. 
As long as $t_L \gg t_J$, the structure formation 
is not affected by the Laplacian instability \cite{Izumi,Mukoreview}. 
In the regime where the wavenumber $k$ is larger than 
$M_3$ and $M_2$ the first two terms on the r.h.s. of 
Eq.~(\ref{dispersion}) dominate over $-k^2$, so the 
Laplacian instability is absent.

In order to recover the behavior close to GR at low energy, 
we require that $\lambda$ is sufficiently close to 1. 
If we expand the original action (\ref{Horava1}) up to $n$-th order ($n>2$) 
in perturbations,  
the $n$-th order Lagrangian density contains the terms with 
negative powers $(\lambda-1)^{-(n-1)}$ (which blows up for 
$\lambda$ close to 1).
For larger $n$ the divergence of these terms in the limit $\lambda \to 1$ 
gets worse, so the perturbative expansion breaks down.
This is the strong coupling problem of the projectable version 
of Ho\v{r}ava-Lifshitz gravity pointed out in Refs.~\cite{strong,strong2}.

\subsection{Non-projectable Ho\v{r}ava-Lifshitz gravity}

The non-projectable version of Ho\v{r}ava-Lifshitz gravity \cite{Sergey}
corresponds to the case where the lapse $N$ depends on 
both $t$ and $x^i$. Since the acceleration $a_i=\nabla_i \ln N$ 
does not vanish, the theory is described by 
the action 
\be
S=\int d^4 x \sqrt{-g} \left[ L(K,{\cal S},{\cal R},{\cal Z},
{\cal Z}_1,{\cal Z}_2\,\alpha_1, \cdots, \alpha_5)
+L_m \right]\,,
\label{Horava2}
\ee
where 
\be
L =
\frac{M_{\rm pl}^2}{2} \left[ {\cal S}-\lambda K^2
+{\cal R}+\eta_1 \alpha_1
-M_{\rm pl}^{-2} \left( g_2 {\cal R}^2+g_3 {\cal Z}+
\eta_2 \alpha_2+\eta_3 \alpha_3\right)
-M_{\rm pl}^{-4}\left( g_4 {\cal Z}_1+g_5 {\cal Z}_2+
\eta_4 \alpha_4+\eta_5 \alpha_5 \right) \right]\,.
\ee

At the background level, the quantities 
${\cal R}, {\cal Z}, {\cal Z}_1, {\cal Z}_2$ and 
$\alpha_i$ ($i=1,2,\cdots,5$) vanish. 
The important difference from the projectable Ho\v{r}ava-Lifshitz 
gravity is that, in the non-projectable version, 
there is the equation of motion derived 
by the variation of $\delta N$. 
In the presence of matter with energy density $\rho_m$ and 
pressure $P_m$, variation of the action (\ref{Horava2})
with respect to $\delta N$ leads to the same form as
Eq.~(\ref{backma1}) and hence
\be
\frac32 (3\lambda-1)M_{\rm pl}^2H^2
=\rho_m\,.
\label{Hoha}
\ee
Unlike Eq.~(\ref{Hoback1}), Eq.~(\ref{Hoha}) does not 
contain the term $C/a^3$. 
The momentum constraint and the matter equation of motion 
are the same as Eqs.~(\ref{Hoback2}) and (\ref{Hoback3}), 
respectively.

In the non-projectable Ho\v{r}ava-Lifshitz gravity it is inconsistent to choose 
the gauge $\delta N(t,x^i)=0$ because the gauge transformation 
(\ref{gauge1}) involves the scalar $\dot{f}$ depending on $t$ alone. 
The temporal gauge transformation is not fixed in the following discussion.
For the discussion of perturbations we do not take into account 
the contribution of matter, but it is straightforward to do so.
Since ${\cal W}=(3\lambda-1)M_{\rm pl}^2H$, 
$L_{,\alpha_1}=M_{\rm pl}^2 \eta_1/2$, 
$L_{,\alpha_2}=-\eta_2/2$, $L_{,\alpha_3}=-\eta_3/2$, 
$L_{,\alpha_4}=-\eta_4/(2M_{\rm pl}^2)$, and 
 $L_{,\alpha_5}=-\eta_5/(2M_{\rm pl}^2)$, 
Eqs.~(\ref{Hami}), (\ref{momen}) and (\ref{zeta2}) 
reduce, respectively, to 
 \ba
 \hspace{-1.2cm}& &
 (3\lambda-1) M_{\rm pl}^2 H
 (3\dot{\zeta}-3H \delta N-\Delta \psi)-2M_{\rm pl}^2 \Delta \zeta
 -M_{\rm pl}^2 \eta_1 \Delta \delta N+\eta_2 \Delta^2 \delta N
 +2\eta_3 \Delta^2 \zeta+\frac{\eta_4}{M_{\rm pl}^2}  \Delta^3 \delta N
 +\frac{2\eta_5}{M_{\rm pl}^2} \Delta^3 \zeta=0\,,
 \label{Hoper1} \\
 \hspace{-1.2cm} & &
 (3\lambda-1)(\dot{\zeta}-H \delta N)
 -(\lambda-1)\Delta \psi=0\,,
 \label{Hoper2} \\
  \hspace{-1.2cm}& &
 \Delta (\dot{\psi}+H\psi+\delta N+\zeta)
 +\frac{\Delta^2 \zeta}{M_2^2}
 -\frac{\Delta^3 \zeta}{M_3^4}
 -\frac{\eta_3}{M_{\rm pl}^2} \Delta^2 \delta N
 -\frac{\eta_5}{M_{\rm pl}^4} \Delta^3 \delta N=0\,,
 \label{Hoper3}
 \ea
where $M_2$ and $M_3$ are defined by Eq.~(\ref{M23}). 
These equations match with Eqs.~(19), (21), and (27) of 
Ref.~\cite{Urakawa}, respectively, in the 
absence of matter perturbations (after the replacement 
$\delta N \to \phi$, $\psi \to a^2 \beta$, $\zeta \to -\psi$, 
and $\eta_1 \to \eta$).
 
Let us discuss the stability of perturbations on the Minkowski 
background ($a=1$ and $H=0$). 
In the IR regime we can ignore the spatial derivatives 
higher than two in the perturbation equations of motion.
{}From Eqs.~(\ref{Hoper1}) and (\ref{Hoper2}) we 
obtain 
\be
\delta N=-\frac{2}{\eta_1} \zeta\,,\qquad
\Delta \psi=\frac{3\lambda-1}{\lambda-1} \dot{\zeta}\,.
\label{delNpsi}
\ee
Substituting these relations into Eq.~(\ref{Hoper3}), it follows that 
\be
\frac{3\lambda-1}{\lambda-1} \ddot{\zeta}
-\frac{2-\eta_1}{\eta_1} \Delta \zeta=0\,.
\ee
The Lagrangian density leading to this equation of motion can be 
derived by substituting the relations (\ref{delNpsi}) into (\ref{L2exp}), as
\be
{\cal L}_2=M_{\rm pl}^2 \frac{3\lambda-1}{\lambda-1} 
\left[ \dot{\zeta}^2-c_s^2 (\partial \zeta)^2 \right]\,,
\label{lagho2}
\ee
where 
\be
c_s^2=\frac{\lambda-1}{3\lambda-1} 
\frac{2-\eta_1}{\eta_1}\,.
\ee
There is a viable parameter space in which both the scalar ghost and 
the Laplacian instability are absent \cite{Sergey}:
\be
\frac{3\lambda-1}{\lambda-1}>0\,,\qquad
0<\eta_1<2\,.
\ee
This is in contrast with the projectable Ho\v{r}ava-Lifshitz gravity 
in which the scalar perturbation in the IR regime is unstable
in the absence of ghosts.
Moreover, the strong-coupling problem in the original Ho\v{r}ava-Lifshitz 
gravity can be alleviated by the presence of non-vanishing 
acceleration terms \cite{Sergey,Blas}.

In order to discuss the cosmology during the radiation and matter 
eras in the non-projectable Ho\v{r}ava-Lifshitz gravity, 
we need to add the contribution of 
matter perturbations to Eqs.~(\ref{Hoper1})-(\ref{Hoper3}). Although 
the temporal gauge transformation is not fixed, one can study
the evolution of cosmological perturbations by considering 
some gauge-invariant variables, 
say $\zeta_{g} \equiv \zeta-H \delta \rho_m/\rho_m$, 
where $\delta \rho_m$ is the matter density perturbation. 
The dynamical evolution of such gauge-invariant variables
has been investigated in detail in Ref.~\cite{Urakawa}.

\section{Conclusions}
\label{consec} 

In this paper we have reviewed the EFT approach to modified gravity 
based on the expansion in terms of cosmological perturbations on 
the flat FLRW background. This approach is powerful enough to 
deal with a wide variety of modified gravitational theories in 
a systematic and unified way. 
Our starting point is the general action (\ref{action}) that 
depends on geometric scalar quantities 
constructed in the 3+1 decomposition of space-time. 
The expansion of this action up to second order in scalar 
metric perturbations allows us to identify the propagating 
scalar degree of freedom.

In addition to the lapse function and several geometric 
scalars arising from the extrinsic and intrinsic curvatures, 
we have taken into account spatial derivatives higher than 
second order such as those given in Eqs.~(\ref{Zdef}) and 
(\ref{aldef}). This generalizes the analysis of Ref.~\cite{Piazza} 
in such a way that our formalism can be applied not only 
to Horndeski theory but also to Ho\v{r}ava-Lifshitz gravity. 
In Sec.~\ref{concretesec} we briefly reviewed both Horndeski 
theory and Ho\v{r}ava-Lifshitz gravity to show explicit relations 
with the EFT approach to modified gravity.

In Sec.~\ref{persec} we expanded the action (\ref{action}) up to 
second order in perturbations for the metric (\ref{permet}) 
with the spatial gauge fixing (\ref{E0}). 
The resulting first-order and second-order 
actions give rise to the background equations (\ref{back1})-(\ref{back2}) 
and the linear perturbation equations (\ref{Hami})-(\ref{zeta}), 
respectively. Under the conditions (\ref{nospa}) the equations of motion 
for perturbations are of second order, in which case 
the second-order Lagrangian density is simply given 
by Eq.~(\ref{perlag}).

In Sec.~\ref{apsec1} we applied our general EFT formalism 
to Horndeski and GLPV theories. 
In these theories the linear perturbation equations of motion 
are of second order on the isotropic cosmological background. 
We provided general formulas for the primordial power spectra of 
scalar and tensor perturbations generated during inflation driven by 
a scalar degree of freedom present in such theories.
We also studied the application to dark energy by taking into 
account an additional matter scalar field and showed that 
Horndeski and GLPV theories can be distinguished from 
each other by the scalar propagation speeds $c_s$. 
In particular the covariantized Galileon (a class of GLPV theories)
is plagued by the Laplacian instability during the matter era 
for late-time tracking solutions, whereas this is not the case
for the covariant Galileon (a class of Horndeski theories).

In Sec.~\ref{apsec2} we discussed the stability of both projectable 
and non-projectable versions of Ho\v{r}ava-Lifshitz gravity 
by employing our general EFT formalism.
In the projectable version the lapse $N$ is a function of the 
time $t$ alone, in which case there are no Hamiltonian constraints 
both for the background and the perturbations. 
In the IR regime the Laplacian instability is present when 
the ghost is absent. 
In the non-projectable version the lapse depends on both space 
and time, so there is freedom to introduce an
acceleration vector $a_i=\nabla_i \ln N$. 
We reproduced the linear perturbation equations on both 
FLRW and Minkowski backgrounds already derived in the 
literature and showed the existence of a 
viable parameter space free from ghosts and instabilities.

We expect that our general EFT formalism
will be useful for the constructions of viable inflation/dark energy models 
as well as quantum gravity. 
This approach is also useful to quantify the non-Gaussianities generated 
during inflation \cite{Cheung,WMAP9,Plancknon} and to parametrize 
the dark energy equation of state 
and density perturbations associated with CMB and large-scale 
structures \cite{Silve2,Piazza3}.
We hope that we will be able to approach the origins of 
inflation/dark energy and to construct renormalizable quantum gravity 
consistent with observations and experiments.

\section*{Acknowledgements}

This article is prepared for Special Issue on modified gravity and effects 
of Lorentz violation in International Journal of Modern Physics D.
This work is supported by the Grant-in-Aid for Scientific 
Research from JSPS (Nos.~24$\cdot$6770 (RK), 24540286 (ST)) 
and by the cooperation programs of Tokyo University of Science 
and CSIC. 


\end{document}